\documentclass[%
 reprint,
 amsmath,amssymb,
 aps,
]{revtex4-2}

\usepackage{graphicx}% Include figure files
\usepackage{dcolumn}% Align table columns on decimal point
\usepackage{bm}% bold math

\begin{document}

\preprint{APS/123-QED}

\title{Geometric Optimization of the MATHUSLA Detector }% Force line breaks with \\

%\author{Imran Alkhatib}
% \altaffiliation[Also at ]{Physics Department, University of Toronto.}%Lines break automatically or can be forced with \\

\author{Imran Alkhatib}%
 \email{alkhatib@physics.utoronto.ca}
\affiliation{%
 Department of Physics, University of Toronto, Toronto Ontario, Canada M5S 1A7\\
}%


%\affiliation{
%\affiliation{
%\affiliation{%



\begin{abstract}
MATHUSLA is a proposed displaced vertex detector for neutral long-lived particle decays. It was
proposed with general specifications of the size of its decay volume and its location. In this study,
different simplified models containing LLPs are investigated using Monte Carlo event generators, and
LLP decay probability maps are generated. Specific optimal configurations for the detector are found
for each model according to available land around the CMS detector. We demonstrate that the placement and dimensions of a proposed 10000 ${m}^2$
 engineering benchmark can be modified so that an improvement in acceptance (up to 12\% more LLP decays) is observed. Also, it is found that the engineering benchmark would observe about 80\% of the number of LLP decays that the earlier MATHUSLA200 physics sensitivity benchmark with four times the area would observe.

\end{abstract}

%\begin{description}
%\end{description}

\maketitle

%\tableofcontents

%\section{\label{sec:level1}First-level heading:\protect\\ The line
%break was forced \lowercase{via} \textbackslash\textbackslash}
\section{\label{sec:level1}Introduction\protect\\ }

The \textbf{MA}ssive \textbf{T}iming \textbf{H}odoscope for \textbf{U}ltra-\textbf{S}table neutra\textbf{L} p\textbf{A}rticles (MATHUSLA) is a
proposed displaced vertex detector at the surface
of the LHC site, designed for the indirect detection of neutral long-lived particles (LLPs). First
proposed in [1], the physical motivation for MATHUSLA includes the existence of neutral LLPs
in many beyond standard model (BSM) theories,
notably various models that address the Hierarchy
Problem, in addition to other BSM models that are
discussed in detail in [2,3]. Neutral uncolored particles produced at the LHC can only be detected
as reconstructed displaced vertices (DVs), or as
missing transverse energy (MET). MET signals
usually have many backgrounds, such as neutrino
MET signals, so a DV detector may have an increased sensitivity to LLPs. In addition, MET signals from the main LHC detectors could be supplemented by DV measurements for a more comprehensive characterization of a neutral particle.

Theoretically, a wide range of particle
lifetimes are possible. However, signatures of the
well understood phenomenon of Big Bang Nucleosynthesis (BBN), which happened about 1 second after the big bang, indicate that any particle
(under some conditions) should have decayed before BBN [1]. Otherwise, cosmological parameters measured today wouldn’t be as they are, unless the particles have a very small energy density,
the particles are stable, or they have a very small
branching ratio into SM hadrons. This sets an upper bound on the lifetimes of some unstable LLPs
produced at the LHC. This upper bound translates
to a decay length of $c\tau  \lesssim 10^7-10^8$
 m [1]. Particles
with lifetimes near this upper bound have a high
probability of escaping the LHC main detectors
before decaying, leaving MET signals that look
identical to MET signals left by stable neutral particles. Thus, it would be instructive to design MATHUSLA so that it has the highest possible acceptance for LLP decays near that limit.

MATHUSLA would be built on the surface above an LHC interaction point (IP). An IP
is about 80 m below the surface of the ground.
Current proposals are made for MATHUSLA to
start 20 m under the surface of the ground, making
the IP 60 m below the bottom surface of MATHUSLA. This 60 m layer of rock would provide
shielding from electrically charged particles and
neutral hadrons from events at the LHC. A detector layer would surround all or part of MATHUSLA’s decay volume to aid in background rejection. The precise detector design and geometry of
MATHUSLA has not yet been determined. The
various design options have features in common
like the ~100 m distance above an IP, with a 20 m
high decay volume, and the closest side being a
maximum of 100 m away horizontally from the
IP. Above the 20 m decay volume would be 5 layers of Resistive Plate Chambers (RPCs) that act as
charged particle trackers, with a separation of 1 m
between each other. The original MATHUSLA200 design, which should be regarded as a
physics sensitivity benchmark, is shown in Figure
1. The 20 m high decay volume was proposed in
the first MATHUSLA paper [1] so that it would
be sensitive to LLPs resulting from Higgs decays
with lifetimes near the BBN limit, for a detector
that has a geometry such that 10\% of produced
LLPs would pass through it. But current proposals
for MATHUSLA call for a 25 m height for the decay volume, in order to achieve a higher acceptance of LLP decays. Both the 20 m and 25 m
high decay volumes were considered in this study,
for a 10000 ${m}^2$
 detector. The only thing left to optimize, then, would be the remaining horizontal
dimensions of MATHUSLA, and its placement
according to available land. Optimizing the placement and the remaining dimensions of MATHUSLA is the aim of this study.

The optimization of the placement and dimensions of MATHUSLA can be considered under several simplified models that are motivated
by BSM theories, as different LLPs with different
production modes would be produced with different distributions at the LHC. A survey done in [4]
highlights four simplified models which include
LLPs, each representing different LLP production
topologies. These are the heavy parent (HP) decay
topology (representative of many supersymmetric
scenarios), the exotic Higgs decay topology (representative of scalar decay to LLPs), the intermediate resonance topology (representative of
gauge-portal Z’ decaying to LLPs), and the exotic
bottom meson decay topology (representative of
heavy neutrino theories) [4]. In this paper, these four LLP production topologies are investigated
via their respective representative models using a
Monte Carlo parton-level event generator (Madgraph5\_aMC@NLO) [6], with showering conducted by Pythia 8. The R-parity violating minimally supersymmetric standard model
(RPVMSSM) is used as a benchmark model for
supersymmetric theories, as is the hidden abelian
Higgs model (HAHM) for Higgs-portal theories,
a new abelian massive gauge field denoted by Z’
(or ZP) for gauge-portal theories, and the plain
standard model (SM) with a right handed neutrino
(RHN) for bottom meson decay into heavy neutral
leptons (HNL) [5].

%%%%%%%%%%%%%%%%%%%%%%%%  FIGURE  1              %%%%%%%%%%%%%%%%%%%%%%%%%%%%%%%%%%%%%%%%%%%%%%%%%

%and Fig.~\ref{fig:epsart}.%
\begin{figure}[t]
\includegraphics[scale=0.8]{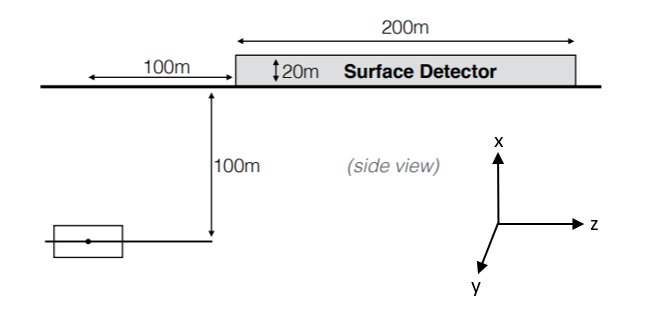}% Here is how to import EPS art
\caption{\label{fig:epsart}  MATHUSLA200 decay volume
(gray shaded), a $(200 m)^2$
 building with its
center along the beam line. It is 100 m above
the IP and 100 m away from the IP in the z-direction. In this coordinate system, the IP is
the origin and the positive x, y, z directions
are shown. Figure from [1]. }
\end{figure}
%%%%%%%%%%%%%%%%%%%%%%%%%%%%%%%%%%%%%%%%%%%%%%%%%%%%%%%%%%%%%%%%%%%%%%%%%%%%%%%%%%%%%

After obtaining a list of LLP momenta
from an event generator, the task would be to find
the probability of the decay of that LLP within a
section of MATHUSLA. This is calculated based
on the simple exponential decay equation evaluated at two lengths $L_1$ and $L_2$ for an LLP with lifetime $\tau$:
%\\
%$P_{decay} (bc\tau, L_1, L_2) = $  \\
%\\
\begin{equation} P_{decay} (bc\tau, L_1, L_2) = e^{\frac{-L_1}{bc\tau}} - e^{\frac{-L_2}{bc\tau}} \approx \frac{L_2-L_1}{bc\tau} ,\end{equation}  %%%%%%%%%
where the approximation holds for the long lifetime limit 
$L_1$, $L_2$ $\ll$ $bc\tau$, $L_1$, $L_2$ are lengths from
the IP to where the LLP enters and leaves a given
volume, respectively, and $b = p/m$ is the boost
of the LLP. Then, the total decay probability
within the entire detector is equal to the sum of the
decay probabilities in each subsection of the detector. Also, it is important to note that under the
same limit, an optimal configuration of MATHUSLA is independent of lifetime, as a change in
lifetime would cause each decay probability within MATHUSLA to change by the same multiplicative factor. Dividing the area surrounding
an LHC IP into squares and using equation (1),
one could generate probability maps for the probability of LLP decays. Then, an optimal configuration for MATHUSLA could be found, maximizing LLP decay acceptance given physical constraints like land availability and assembly requirements.

%\footnote{Automatically placing footnotes into the bibliography requires using BibTeX to compile the bibliography.};
%author-year citation styles place the footnote at the bottom of the text column.
%Note: due to the method used to place footnotes in the bibliography, 
%\emph{you must re-run Bib\TeX\ every time you change any of your document's footnotes}. 

%%%%%%%%%%%%%%%%%%%%%%%%%%%%%%%%%%%%%%%%%%%%%%%%%%%%%%%%%%%%%%%%%%%%%%%%%%%%%
%%%%%%%%%%%%%%%%%%%%%%%%%%%%%%%%%%%%%%%%%%%%%%%%%%%%%%%%%%%%%%%%%%%%%%%%%%%%%%
%%%%%%%%%%%%%%%%%%%%%%%%%%%%%%%%%%%%%%%%%%%%%%%%%%%%%%%%%%%%%%%%%%%%%

\section{Methods}

There are three steps for creating LLP decay probability maps in an area surrounding an
LHC IP, namely, event generation and showering;
event selection or reweighting; and creating the
probability maps using the momenta from the
event generations. Afterwards, an optimization of
MATHUSLA design parameters would be conducted on each probability map. These four steps
of MATHUSLA optimization are discussed in detail in the following subsections.

\subsection{Parton-Level Event Generation and Showering}

For each model of RPVMSSM, HAHM,
ZP, and SM, events were generated at parton level
using Madgraph5\_aMC@NLO v.2.6.5 (MG5)
[6], interfaced with Pythia 8.240. Pythia was used
to conduct the parton showering and hadronization on the MG5 output. All sets of events generated for the different models are listed in Table I.
For the RPVMSSM, a heavy parent (HP) decay
was considered, where a heavy colored parent particle (a squark) decays into quarks and a neutralino, which is taken to be the neutral LLP [4].
Each event is generated in MG5 as squark pair
production from pp collisions, and subsequent decay of both squarks into a neutralino each, with a
dijet signature for each event.

For the HAHM model, gluon fusion and
vector boson fusion (VBF) events were generated
separately. For gluon fusion, events were generated as Higgs boson production from pp collisions
with a monojet signature, followed by Higgs decay into a pair of neutral scalars, which were taken
to be the LLPs. For VBF Higgs production, W
boson fusion, Z boson fusion and photon fusion
events produced a Higgs boson from pp interactions with a dijet signature, followed by Higgs decay to a pair of LLPs. For the ZP model, events
were generated as vector boson Z’ production
from pp collisions with a monojet signature, followed by decay of Z’ to a pair of neutral LLPs.
Depending on the mass of Z’, the production
modes of those LLPs differ. For Z’ masses under
a couple TeV, the Z’ is produced on-shell at the
14 TeV LHC, and the subsequent decay into LLPs
is denoted by heavy resonance (RES) production
[4]. For Z’ masses around 10 TeV, the Z’ is offshell and the LLPs are produced by direct-pair
production (DPP) [4]. For each of the events generated for the HAHM and ZP models, a jet matching procedure was used to prevent double counting between MG5 and Pythia, discussed further in
subsection B of this section.

%%%%%%%%%%%%%%%%%%%%%%%%%%%%         Table I               %%%%%%%%%%%%%%%%%%%%%%%%%%%%%%%%%%%5
%Table~\ref{tab:table1},%

\begin{table*}[t]%The best place to locate the table environment is directly after its first reference in text

\begin{ruledtabular}
\begin{tabular}{lccr}
\textrm{Model}&
\textrm{$m_{LLP}$ (GeV)}&
\textrm{MG5 Input}&
\textrm{Events Generated} \\
\colrule
HAHM, gluon fusion  & 5, 15, 30, 50  & p p $>$ h, h $>$ hs hs  & 2 $\times 10^6$ \quad per \quad $m_{LLP}$\\
 & & p p $>$ h j, h $>$ hs hs & \\
\\
HAHM, VBF  & 5, 15, 30, 50  & p p $>$ h j j \$\$w+ w- z a, & 1 $\times 10^6$ \quad per \quad $m_{LLP}$\\
 & & h $>$ zp zp QCD=0 & \\
\\
\\
ZP, $m_{ZP}$ = 400 GeV & 20, 100, 200  & p p $>$ zp, zp $>$ x2 x2$\sim$ & 2 $\times 10^6$ \quad per \quad $m_{LLP}$\\
ZP, $m_{ZP}$ = 2 TeV   & 100, 500, 1000  &  p p $>$ zp j, zp $>$ x2 x2$\sim$ & 2 $\times 10^6$ \quad per \quad $m_{LLP}$\\
ZP, $m_{ZP}$ = 10 TeV   & 10, 100, 1000  &  & 2 $\times 10^6$ \quad per \quad $m_{LLP}$\\
\\
\\
RPVMSSM, $m_y$ = 500 GeV & 50, 125, 400  & p p $>$ su6 su6$\sim$, & 1 $\times 10^6$ \quad per \quad $m_{LLP}$\\
RPVMSSM, $m_y$ = 2 TeV   & 200, 500, 1200  & su6 $>$ n4 j, su6$\sim$ $>$ n4 j  & 1 $\times 10^6$ \quad per \quad $m_{LLP}$\\
\\
\\
pp $\rightarrow$ B $\rightarrow$ RHN  & 0.1, 4 & p p $>$ b b$\sim$  & 2 $\times 10^7$\\
pp $\rightarrow$ B $\rightarrow$ scalar & 0.1, 4  &  & \\

\end{tabular}
\end{ruledtabular}

\caption{\label{tab:table1}%
 Events generated by MG5 and showered by Pythia. Different mediator and LLP masses used
are shown, in addition to number of events generated and the MG5 input used to generate those events.
For processes where jet matching was used, about half the events were rejected so twice as many events
were generated compared to other non-matched samples. For the RPVMSSM model, y denotes the
heavy parent particle. 
}
\end{table*}
%%%%%%%%%%%%%%%%%%%%%%%%%%%%%%%%%%%%%%%%%%%%%%%%%%%%%%%%%%%%%%%%%%%%%%%%%%%%%%%%%%%%%%%%%%%%%%%%%%%%%%%%%%%

Finally, for the heavy right-handed neutrino model, a bottom and an anti-bottom quark
are pair produced from p p collisions, followed by
hadronization into bottom mesons. Bottom meson
decays into right-handed neutrinos were done outside of MG5 and Pythia, according to decay
modes provided by [5]. Bottom mesons $ B^{\pm}$, $B^0$ and $B_s$ were extracted from events and decayed into
right-handed neutrinos. Bottom meson decays
into a complex scalar were also done for comparison ($B^{+} \rightarrow K^0 \phi $ and $B^0 \rightarrow K^+ \phi$) [10]. The decay
modes used along with the branching ratios are
shown in Table II. A random direction was chosen
in the frame of the mesons, and a momentum was
calculated based on 2-body or 3-body decay kinematics. Then, that momentum was boosted to the
lab frame. However, in this case, each event was
reweighed according to “Fixed Order + Next-toLeading Log” (FONLL) predictions for bottom
meson production at the LHC [7], provided by [8].
This reweighting of events is described in subsection B of this section.

\subsection{Jet Matching and FONLL}

For the HAHM and ZP model events, jet
matching was used to prevent overlapping between phase-space calculations from the parton
level event generator and the hadronization software, since there are intermediate jets in these
events. A kt jet matching procedure was used,
where a minimum value (xqcut) was set for the kt
jet measure allowed for an event to be accepted.
The value of xqcut was optimized by finding the
value of xqcut such that jet transverse momentum (PT) distributions are smooth and do not change
around that value of xqcut. For HAHM model
gluon fusion events, the optimal value of xqcut
was around $1/6$ of the hard scale, which is the
Higgs mass, so xqcut was set to 20 GeV. It was
also found that for ZP heavy resonance production, the optimal xqcut value was 10\% of the hard
scale (which is the mass of Z’), and for the case of
direct pair production it was 20\% of the hard scale
(which is 2 times the LLP mass).

To account for QCD effects at the scale of
bottom quark production, a reweighting of events
according to PT distributions is warranted. For
this, a FONLL bottom meson PT distribution
spanning the range from PT = 0 to 30 GeV was
generated from [8] in increments of 0.5 GeV.
About 99.5\% of all bottom mesons produced at
the LHC have PT values in that range according
to MG5+Pythia predictions. Another PT distribution was obtained from the bottom mesons that
were generated by MG5+Pythia. Both PT distributions were normalized. Afterwards, for
each generated MG5+Pythia event, the bottom
meson PT was calculated, and if the PT value was
between 0 and 30 GeV, the value of that PT was
rounded up to the nearest 0.5 GeV (only for the
purposes of finding a reweighting factor), and a
reweighting factor was generated by dividing the
value of the FONLL distribution function at that
PT by the value of the MG5+Pythia distribution
function at the same PT. These reweighting factors were used in creating the LLP decay probability maps, where each decay probability resulting from an event is multiplied by the reweighting
factor associated with that event. This is discussed
further in subsection C of this section.

\subsection{Probability map Generation}

To generate probability maps for each
process, a 300 m by 300 m area in the positive y-z plane (Figure 1) was divided into 5 m by 5 m
boxes. These boxes represent cuboids that are 25
m high, and the distance of the cuboids’ bottom
surfaces was set to 60, 70, 80, 90 and 100 m above
the IP in the x-direction to test acceptance at different distances from the IP. Then, for accepted
events of the HAHM and ZP models, in addition
to RPVMSSM events, each LLP momentum was
traced until it traversed the entire 300 m by 300 m
area, and the amount of length spent within each
volume was calculated, so that a value for $P_{decay}$ from equation (1) would be found for each event
in each cuboid. All $P_{decay}$ values found for a cuboid i were added and the sum $P_{tot,i}$ was recorded. If one notes azimuthal symmetry about
the z-axis in addition to $\pm z$  symmetry, then one
can use a trick to appear to increase the generated
LLP sample size so that a smoother probability
map would be produced with minimal computational costs. This means that if every negative
LLP momentum component in our coordinate
system has its sign flipped, so that all LLPs would
pass through one octant of a sphere around the IP,
we’d create an LLP decay probability map that
would look as if we had used a dataset eight times
as large. Then the total number of LLPs that
would decay within a cuboid i would be:
\begin{equation} N_{decay,i} = (\sigma \mathcal{L}) \frac{n_{LLP}}{8 N_{LLP}} P_{tot,i} = (\sigma \mathcal{L}) P_{map,i} ,\end{equation}
where $\sigma$ is the production cross section of the LLP
at the LHC, $\mathcal{L}$ is the total integrated luminosity of
the LHC run, $n_{LLP}$ is the number of LLPs produced
per event and $N_{LLP}$ is the total number of LLPs
from the generated events. Then, each probability
map was generated with values of $P_{map,i}$ and an
LLP lifetime of 1 second, so that the probability
maps could be rescaled for any LLP lifetime by
dividing by the lifetime in seconds. The same procedure for creating heavy right-handed neutrino
decay maps was used, except in this case each $P_{decay}$ value that contributed to the sum $P_{tot,i}$  is
multiplied by the FONLL reweighting factor depending on the PT of the parent bottom meson.

For the HAHM model, VBF and gluon
fusion maps were combined to form one LLP decay probability map for a given LLP mass. The
VBF and gluon fusion maps were added according to 14 TeV LHC Higgs production cross
sections for a 125 GeV Higgs boson [9]. LLPs resulting from bottom meson decays had their probability maps combined according to branching ratios in Table II, and according to the production
cross sections of the bottom mesons. These
branching ratios and cross sections were normalized so that each bottom meson would decay into
a heavy RHN, and the cross sections were simply
relative cross sections (the fraction of bottom mesons produced that were a specific type for example).

\subsection{MATHUSLA Optimization}

After generating probability maps spanning the positive y-z plane for each model, the
probability maps were reflected across the z-axis,
and then again across the y axis so that they cover
the entire y-z plane. Figure 2 shows the available
land around CMS, with an engineering benchmark MATHUSLA design that is 100 m by 100
m. The red line on the boundary of available land
indicates where one side of a rectangular MATHUSLA should be placed, according to engineering requirements.

An optimization ran on each LLP decay
probability map to find the optimal placement and
dimensions that give a 10000 $m^2$ MATHUSLA
the highest acceptance (highest value of $\sum_{i \in configuration} P_{map,i}$) for LLP decays within
the boundaries of available land. For this, distances were divided in increments of 1/3 of a meter. After finding the optimal MATHUSLA configuration, it was compared to the engineering
benchmark in Figure 2, in addition to the MATHUSLA200 design shown in Figure 1, where
MATHUSLA200 is 100 m above the IP and has a
20 m high decay volume. Since the optimization
increment size used was smaller than the increment size of the probability maps (which was 5
m), if a configuration of MATHUSLA included a
fraction $f$  of the area of a 5 m by 5 m square i in
a probability map, then $f \times P_{map,i}$ was added in the
calculation of the total decay probability within
that configuration.

%%%%%%%%%%%%%%%%%%%%%%%%  FIGURE  2              %%%%%%%%%%%%%%%%%%%%%%%%%%%%%%%%%%%%%%%%%%%%%%%%%

%and Fig.~\ref{fig:epsart}.%
\begin{figure}[b]
\includegraphics[scale=0.5]{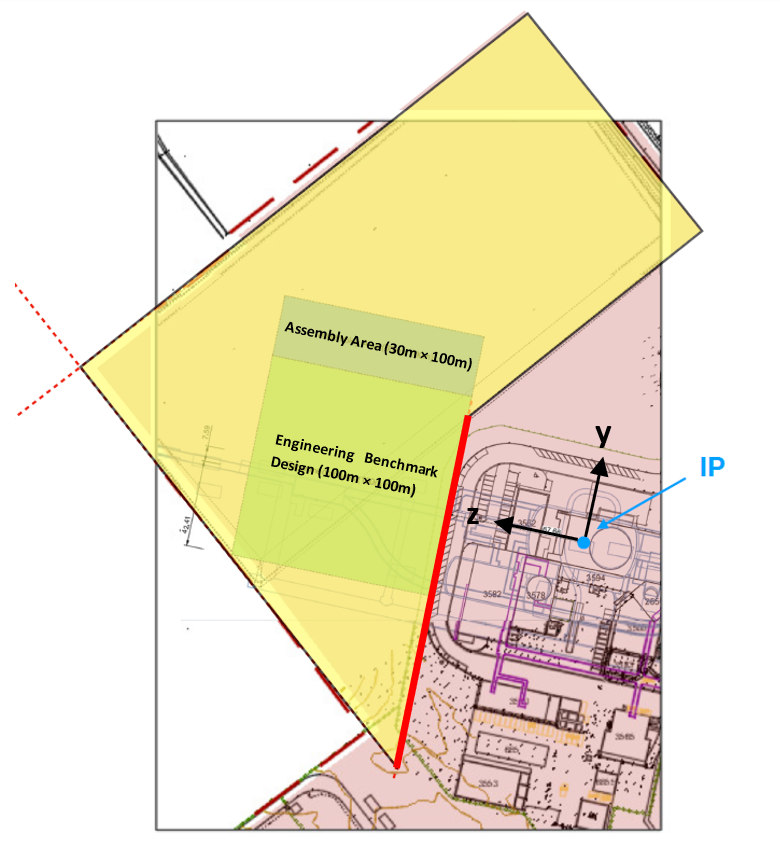}% Here is how to import EPS art
\caption{\label{fig:epsart} Available land around CMS (yellow). The IP is 80 m deep below the ground
and is 67.66 m away from the red shaded part
of the boundary of the available land. The coordinate system of the y-z plane is shown
around the IP, along with the engineering
benchmark design of MATHUSLA and its assembly area.  }
\end{figure}
%%%%%%%%%%%%%%%%%%%%%%%%%%%%%%%%%%%%%%%%%%%%%%%%%%%%%%%%%%%%%%%%%%%%%%%%%%%%%%%%%%%%%

%%%%%%%%%%%%%%%%%%%%%%%%%%%%%%%%%%%%%%%%     Table II         %%%%%%%%%%%%%%%%%%%%%%%%%%%%%%%%%%%%%%

\begin{table}[t]%The best place to locate the table environment is directly after its first reference in text

\begin{ruledtabular}
\begin{tabular}{lcc}
\textrm{Process}&
\textrm{$BR(m_N=0.1 GeV)$}&
\textrm{$BR(m_N=4.0 GeV)$} \\
\colrule
 $B^+ \rightarrow e+N$ & $<10^{-6}$  &  $1.3\times10^{-4}$\\
 $B^+ \rightarrow D^0+e+N$ & $2.8\times10^{-2}$  &  $0$\\
 $B^+ \rightarrow D^{0*} +e+N$ &  $6.5\times10^{-2}$ &  $0$\\
 $B^0 \rightarrow D^+ +e+N$ & $2.5\times10^{-2}$  &  $0$\\
 $B^0 \rightarrow D^{+*} +e+N$ & $5.7\times10^{-2}$  &  $0$\\
 $B^0 \rightarrow \pi^+ +e+N$ & $1.2\times10^{-4}$  &  $8.4\times10^{-6}$\\
 $B^0 \rightarrow \rho^+ +e+N$ & $3.3\times10^{-4}$  &  $1.3\times10^{-6}$\\
 $B_s \rightarrow D_s+e+N$ & $2.2\times10^{-2}$  &  $0$\\
 $B_s \rightarrow D_s^*+e+N$ & $5.0\times10^{-2}$  &  $0$\\

\end{tabular}
\end{ruledtabular}

\caption{\label{tab:table2}%
Bottom meson decays into right-handed neutrinos that were considered.
Branching ratios for a 0.1 GeV neutrino and a
4.0 GeV neutrino are provided by [5]. 
}
\end{table}

%%%%%%%%%%%%%%%%%%%%%%%%%%%%%%%%%%%%%%%%%%%%%%%%%%%%%%%%%%%%%%%%%%%%%%%%%%%%%%%%%%%%

\section{Results}

As described in subsection D of the previous section, rectangular configurations of MATHUSLA were tried and an optimal configuration
was identified if it maximized the value of $P = \sum_{i \in configuration} P_{map,i}$. The results of the optimization for a detector 60 m above the IP with a
25 m high decay volume are shown in Table III.
Plots of the probability maps are shown in Appendix A. The results of the optimization show minimal improvements in the acceptance of MATHUSLA from the engineering benchmark (up to
12.2\% in the long lifetime limit). The best improvements occurred for the RPVMSSM, especially for a heavy parent particle with mass 2 TeV
and low LLP masses. For the ZP model, the best
improvements were recorded for Z’ masses in the
TeV scale, where improvements were up to 7.0\%.
For the HAHM model, the maximum improvement was recorded for the lowest LLP mass tested
(5 GeV), where improvements were 3.8\%. For
LLPs resulting from bottom meson decays, improvements in acceptance were very small (up to
1.9\%).

\section{Discussion and Conclusions}

The geometric optimization of a 10000
$m^2$
 MATHUSLA showed that improvements to
the acceptance of the engineering benchmark design in the long lifetime limit ($L_1, L_2 \ll bc\tau$) would be minimal. It also showed that the most
considerable improvements to acceptance can result for the lightest of LLPs. Light LLPs may have longer lifetimes due to kinematic suppression of
their decay, and an increase of around 12\% to the
decays of such particles within MATHUSLA may
be useful for no additional costs. In general, for
light LLPs, an increase in the number of decays
within a 10000 $m^2$
 MATHUSLA was found to be
possible if its y-dimension was longer than 100 m,
while its z-dimension was shorter. Furthermore,
this study has not considered the short lifetime
limit. For short-lived particles, it is expected that
decays will occur close to the IP, also favouring a
flatter MATHUSLA in the z-direction.

The engineering benchmark appeared to
be a robust configuration for MATHUSLA in
terms of LLP decay acceptance. The minimal
change in the acceptance of MATHUSLA for
LLP decays in the long lifetime limit depending
on the y-z configuration means that the detector
can be assembled in almost any configuration, as
determined by engineering requirements, as long
as it is placed as far towards the negative y-direction as possible in the available land. Also, the robustness of the engineering benchmark was reflected in the closeness of the total LLP decay
probabilities of the engineering benchmark and
MATHUSLA200. Although the engineering
benchmark covers a quarter of the area covered by
MATHUSLA200, it would observe around 80\%
of the number LLP decays observed within MATHUSLA200, consistent across all models and
LLP masses tested. This is significant since similar acceptance to MATHUSLA200 can be
achieved with a quarter of the area, and commensurably reduced costs.

The results of this study can be applied to
many scenarios of LLPs at the LHC, since multiple simplified models from the survey of LLP theories done in [4] were considered. A wide range
of mediator and LLP masses were also considered. Searches for lighter LLPs may benefit from
a flatter MATHUSLA in the z-direction. But for
LLP production processes considered in this study
(in the long lifetime limit), this improvement is
not expected to exceed ~10\% of the current engineering benchmark. A larger improvement is expected in the short lifetime limit if more of the detector is placed closer to the IP. 
%\\
%\\

%%%%%%%%%%%%%%%%%%%%%%%%%%%%%%%%%%%%%%%%%%%
%%%%%%%%%%%%%%%%%%%%%%%%%%%%%%%%%%%%%%%%%%%
%%%%%%%%%%%%%%%%%%%%%%%%%%%%%%%%%%%%%%%%%%%
%%%%%%%%%%%%%%%%%%%%%%%%%%%%%%%%%%%%%%%%%%%%%%%%%%%%%%%%%%%%%%%%%%%%%%%%%%%%%%%%%%%%%%
%%%%%%%%%%%%%%%%%%%%%%%%%%%%%%%%%%%%%%%%%%%
%%%%%%%%%%%%%%%%%%%%%%%%%%%%%%%%%%%%%%%%%%%
%%%%%%%%%%%%%%%%%%%%%%%%%%%%%%%%%%%%%%%%%%%

%%%%%%%%%%%%%%%%%%%%%%%%%%%%%%%%%%%%%%%%%%%%%%%%%%%%%%%%%%%%%%%%%%%%%%%%%%%%%%%%%%%%%%%%%%%%%%%%%%%%%%%%%%%

%%%%%%%%%%%%%%%%%%%%%%%%%%%%%%%%%%%%%%%%%%%
%%%%%%%%%%%%%%%%%%%%%%%%%%%%%%%%%%%%%%%%%%%
%%%%%%%%%%%%%%%%%%%%%%%%%%%%%%%%%%%%%%%%%%%
\begin{acknowledgments}
We thank Professor David Curtin of the
Department of Physics at the University of Toronto for the guidance and advice he provided for
the completion of this study. We also thank PhD
student Jared Barron of the Department of Physics at the University of Toronto for providing the
optimized jet matching parameters xqcut 
\dots.
%\\
%\\
%\\
%\\
%\\
%\\
%\\
%\\
%\\

\end{acknowledgments}
%%%%%%%%%%%%%%%%%%%%%%%%%%%%%%%%%%%%%%%%%%%
%%%%%%%%%%%%%%%%%%%%%%%%%%%%%%%%%%%%%%%%%%%
%%%%%%%%%%%%%%%%%%%%%%%%%%%%%%%%%%%%%%%%%%%

%%%%%%%%%%%%%%%%%%%%%%%%%%%%%%%%%%%%     Table III      %%%%%%%%%%%%%%%%%%%%%%%%%%%%%%%%%%%%%%%%%%%%%%%%
\begin{table*}[t]%The best place to locate the table environment is directly after its first reference in text

\begin{ruledtabular}
\begin{tabular}{lcccccccc}
\textrm{Model}&
\textrm{$m_{LLP}$ (GeV)}&
\textrm{$y_1$ (m)}&
\textrm{$z_1$ (m)}&
\textrm{$L_y$ (m)}&
\textrm{$L_z$ (m)}&
\textrm{$\frac{P_{opt}}{P_{eng}}$}&
\textrm{$\frac{P_{opt}}{P_{MAT200}}$}&
\textrm{$\frac{P_{eng}}{P_{MAT200}}$} \\
\colrule

 HAHM & 5  &  -73.67 & 67.67 & 146.33 & 68.33 & 1.038 & 0.856 &  0.825   \\
 HAHM & 15  & -69.33  & 67.67 & 136.33 & 73.33 & 1.030 & 0.848 &  0.823   \\
 HAHM & 30  & -67.33  & 67.67 & 132.00 & 75.67 & 1.018 & 0.834 & 0.819    \\
 HAHM & 50  &  -51.67 & 67.67 & 106.00 & 94.33  & 1.002 & 0.812  &   0.810  \\
\\
 ZP, $m_{ZP} = 0.4 \quad TeV$ & 20  & -69.33  & 67.67  & 136.33  & 73.33  & 1.037 & 0.853 &  0.823   \\
 ZP, $m_{ZP} =0.4 \quad TeV$ & 100  & -56.33  & 67.67 & 112.33 & 89.00 & 1.007 & 0.823 &  0.817   \\
 ZP, $m_{ZP} =0.4 \quad TeV$ & 200  &  -30.33 & 67.67 & 83.33 & 120.00 & 1.022 & 0.806 &   0.789  \\
 ZP, $m_{ZP} = 2 \quad TeV$  & 100  & -73.67  & 67.67  & 146.33 & 68.33 & 1.052 & 0.870 &  0.827   \\
 ZP, $m_{ZP} = 2 \quad TeV$ & 500  & -78.00  & 67.67 & 158.67 & 63.00 & 1.069 & 0.887 &  0.830   \\
 ZP, $m_{ZP} = 2 \quad TeV$ & 1000  & -30.33  & 67.67 & 83.33 & 120.00 & 1.020 & 0.805 &  0.789   \\
 ZP, $m_{ZP} = 10 \quad TeV$ & 10  & -78.00  & 67.67 & 158.67  & 63.00 & 1.068 & 0.890 &   0.833  \\
 ZP, $m_{ZP} = 10 \quad TeV$ & 100  & -75.33  & 67.67 & 150.67 & 66.33 & 1.067 & 0.889 &  0.833   \\
 ZP, $m_{ZP} = 10 \quad TeV$ & 1000  & -78.00  & 67.67 & 158.67 & 63.00  & 1.070 & 0.891 &  0.833   \\
\\
 RPVMSSM, $m_y = 0.5 \quad TeV$ & 50  & -79.33  & 67.67 & 163.00 & 61.33 & 1.080 & 0.898 &  0.831   \\
 RPVMSSM, $m_y = 0.5 \quad TeV$  & 125  & -78.00  & 67.67 & 158.67 & 63.00 & 1.069 & 0.889 &  0.832   \\
 RPVMSSM, $m_y = 0.5 \quad TeV$ & 400  & -55.00  & 67.67 & 110.67 & 90.33 & 1.005  & 0.822 & 0.818     \\
 RPVMSSM, $m_y = 2 \quad TeV$ & 200  &  -84.33  & 67.67 & 180.67 & 55.33  & 1.122  & 0.937 &  0.835   \\
 RPVMSSM, $m_y = 2 \quad TeV$ & 500  & -82.67  & 67.67 & 174.33 & 57.33 & 1.121 & 0.937 &  0.836   \\
 RPVMSSM, $m_y = 2 \quad TeV$ & 1200  & -80.33   & 67.67 & 165.67 & 60.33 & 1.097 & 0.916 &   0.835  \\
\\
 p p $\rightarrow$ B $\rightarrow$ RHN & 0.1  & -67.33  & 67.67 & 132.00 & 75.67 & 1.013  & 0.830 & 0.819    \\
 p p $\rightarrow$ B $\rightarrow$ RHN & 4.0  & -46.00  & 67.67 & 98.67 & 101.33 & 1.002 & 0.809 &  0.807    \\
 p p $\rightarrow$ B $\rightarrow$ scalar & 0.1  & -69.33  & 67.67 & 136.33 & 73.33 & 1.019 & 0.835 & 0.819    \\
 p p $\rightarrow$ B $\rightarrow$ scalar & 4.0  &  -46.00 & 67.67 & 98.67 & 101.33  & 1.002 & 0.809 & 0.807    \\

\end{tabular}
\end{ruledtabular}

\caption{\label{tab:table3}%
Summary of the MATHUSLA optimization results for a detector 60 m above the IP and a 25
m high decay volume. $(y_1, z_1)$ is the position of the lower left corner of the optimal MATHUSLA
configuration, in a coordinate system where the positive z direction points upwards, and the positive y
direction points to the right. $(L_y, L_z)$ are the $(y, z)$ dimensions of the optimal MATHUSLA configuration, respectively. $P_{opt} = \sum_{i \in optimal \quad configuration} P_{map,i}$, $P_{eng} = \sum_{i \in engineering \quad benchmark} P_{map,i}$, and $P_{MAT200} = \sum_{i \in MATHUSLA200} P_{map,i}$, where the MATHUSLA200 design is shown in Figure 1. 
}

\end{table*}
%%%%%%%%%%%%%%%%%%%%%%%%%%%%%%%%%%%%%%%%%%%%%%%%%%%%%%%%%%%%%%%%%%%%%%%%%%%%%%%%%%%%%%%%%%%%%%%%%%%

\nocite{*}

%\bibliography{}
\bibliography{bib}% Produces the bibliography via BibTeX.

\providecommand{\noopsort}[1]{}\providecommand{\singleletter}[1]{#1}%
\begin{thebibliography}{10}%
\makeatletter
\providecommand \@ifxundefined [1]{%
 \@ifx{#1\undefined}
}%
\providecommand \@ifnum [1]{%
 \ifnum #1\expandafter \@firstoftwo
 \else \expandafter \@secondoftwo
 \fi
}%
\providecommand \@ifx [1]{%
 \ifx #1\expandafter \@firstoftwo
 \else \expandafter \@secondoftwo
 \fi
}%
\providecommand \natexlab [1]{#1}%
\providecommand \enquote  [1]{``#1''}%
\providecommand \bibnamefont  [1]{#1}%
\providecommand \bibfnamefont [1]{#1}%
\providecommand \citenamefont [1]{#1}%
\providecommand \href@noop [0]{\@secondoftwo}%
\providecommand \href [0]{\begingroup \@sanitize@url \@href}%
\providecommand \@href[1]{\@@startlink{#1}\@@href}%
\providecommand \@@href[1]{\endgroup#1\@@endlink}%
\providecommand \@sanitize@url [0]{\catcode `\\12\catcode `\$12\catcode
  `\&12\catcode `\#12\catcode `\^12\catcode `\_12\catcode `\%12\relax}%
\providecommand \@@startlink[1]{}%
\providecommand \@@endlink[0]{}%
\providecommand \url  [0]{\begingroup\@sanitize@url \@url }%
\providecommand \@url [1]{\endgroup\@href {#1}{\urlprefix }}%
\providecommand \urlprefix  [0]{URL }%
\providecommand \Eprint [0]{\href }%
\providecommand \doibase [0]{https://doi.org/}%
\providecommand \selectlanguage [0]{\@gobble}%
\providecommand \bibinfo  [0]{\@secondoftwo}%
\providecommand \bibfield  [0]{\@secondoftwo}%
\providecommand \translation [1]{[#1]}%
\providecommand \BibitemOpen [0]{}%
\providecommand \bibitemStop [0]{}%
\providecommand \bibitemNoStop [0]{.\EOS\space}%
\providecommand \EOS [0]{\spacefactor3000\relax}%
\providecommand \BibitemShut  [1]{\csname bibitem#1\endcsname}%
\let\auto@bib@innerbib\@empty
%</preamble>
\bibitem [{\citenamefont {Chou}\ \emph {et~al.}(2017)\citenamefont {Chou},
  \citenamefont {Curtin},\ and\ \citenamefont {Lubatti}}]{Chou_2017}%
  \BibitemOpen
  \bibfield  {author} {\bibinfo {author} {\bibfnamefont {J.~P.}\ \bibnamefont
  {Chou}}, \bibinfo {author} {\bibfnamefont {D.}~\bibnamefont {Curtin}},\ and\
  \bibinfo {author} {\bibfnamefont {H.}~\bibnamefont {Lubatti}},\ }\bibfield
  {title} {\bibinfo {title} {New detectors to explore the lifetime frontier},\
  }\href {https://doi.org/10.1016/j.physletb.2017.01.043} {\bibfield  {journal}
  {\bibinfo  {journal} {Physics Letters B}\ }\textbf {\bibinfo {volume}
  {767}},\ \bibinfo {pages} {29–36} (\bibinfo {year} {2017})},\ \Eprint
  {https://arxiv.org/abs/1606.06298} {arXiv:1606.06298 [hep-ph]} \BibitemShut
  {NoStop}%
\bibitem [{\citenamefont {et~al.}()}]{2}%
  \BibitemOpen
  \bibfield  {author} {\bibinfo {author} {\bibfnamefont {C.~A.}\ \bibnamefont
  {et~al.}} (\bibinfo {collaboration} {MATHUSLA Collaboration}),\ }\bibfield
  {title} {\bibinfo {title} {A letter of intent for mathusla: a dedicated
  displaced vertex detector above atlas or cms.},\ }\href@noop {} {\ }\Eprint
  {https://arxiv.org/abs/arXiv:1811.00927 [physics.ins-det]} {arXiv:1811.00927
  [physics.ins-det]} \BibitemShut {NoStop}%
\bibitem [{\citenamefont {Curtin}\ \emph {et~al.}(2019)\citenamefont {Curtin},
  \citenamefont {Drewes}, \citenamefont {McCullough}, \citenamefont {Meade},
  \citenamefont {Mohapatra}, \citenamefont {Shelton},\ and\ \citenamefont
  {Shuve}}]{3}%
  \BibitemOpen
  \bibfield  {author} {\bibinfo {author} {\bibfnamefont {D.}~\bibnamefont
  {Curtin}}, \bibinfo {author} {\bibfnamefont {M.}~\bibnamefont {Drewes}},
  \bibinfo {author} {\bibfnamefont {M.}~\bibnamefont {McCullough}}, \bibinfo
  {author} {\bibfnamefont {P.}~\bibnamefont {Meade}}, \bibinfo {author}
  {\bibfnamefont {R.}~\bibnamefont {Mohapatra}}, \bibinfo {author}
  {\bibfnamefont {J.}~\bibnamefont {Shelton}},\ and\ \bibinfo {author}
  {\bibfnamefont {B.}~\bibnamefont {Shuve}},\ }\bibfield  {title} {\bibinfo
  {title} {Long-lived particles at the energy frontier: The mathusla physics
  case.},\ }\href@noop {} {\bibfield  {journal} {\bibinfo  {journal} {Reports
  on Progress in Physics.}\ } (\bibinfo {year} {2019})},\ \Eprint
  {https://arxiv.org/abs/arXiv:1806.07396v2 [hep-ph]} {arXiv:1806.07396v2
  [hep-ph]} \BibitemShut {NoStop}%
\bibitem [{\citenamefont {et~al.}(2017)}]{4}%
  \BibitemOpen
  \bibfield  {author} {\bibinfo {author} {\bibfnamefont {J.~A.}\ \bibnamefont
  {et~al.}},\ }\bibfield  {title} {\bibinfo {title} {Searching for long-lived
  particles beyond the standard model at the large hadron collider.},\
  }\href@noop {} {\bibfield  {journal} {\bibinfo  {journal} {Physics Letters
  B}\ }\textbf {\bibinfo {volume} {767}},\ \bibinfo {pages} {29 to 36}
  (\bibinfo {year} {2017})},\ \Eprint {https://arxiv.org/abs/arXiv:1903.04497
  [hep-ex]} {arXiv:1903.04497 [hep-ex]} \BibitemShut {NoStop}%
\bibitem [{\citenamefont {Bondarenko}\ \emph {et~al.}(2018)\citenamefont
  {Bondarenko}, \citenamefont {Boyarsky}, \citenamefont {Gorbunov},\ and\
  \citenamefont {Ruchayskiy}}]{Bondarenko_2018}%
  \BibitemOpen
  \bibfield  {author} {\bibinfo {author} {\bibfnamefont {K.}~\bibnamefont
  {Bondarenko}}, \bibinfo {author} {\bibfnamefont {A.}~\bibnamefont
  {Boyarsky}}, \bibinfo {author} {\bibfnamefont {D.}~\bibnamefont {Gorbunov}},\
  and\ \bibinfo {author} {\bibfnamefont {O.}~\bibnamefont {Ruchayskiy}},\
  }\bibfield  {title} {\bibinfo {title} {Phenomenology of gev-scale heavy
  neutral leptons},\ }\bibfield  {journal} {\bibinfo  {journal} {Journal of
  High Energy Physics}\ }\textbf {\bibinfo {volume} {2018}},\ \href
  {https://doi.org/10.1007/jhep11(2018)032} {10.1007/jhep11(2018)032} (\bibinfo
  {year} {2018}),\ \Eprint {https://arxiv.org/abs/1805.08567} {arXiv:1805.08567
  [hep-ph]} \BibitemShut {NoStop}%
\bibitem [{\citenamefont {Alwall}\ \emph {et~al.}(2014)\citenamefont {Alwall},
  \citenamefont {Frederix}, \citenamefont {Frixione}, \citenamefont {Hirschi},
  \citenamefont {Maltoni}, \citenamefont {Mattelaer}, \citenamefont {Shao},
  \citenamefont {Stelzer}, \citenamefont {Torrielli},\ and\ \citenamefont
  {Zaro}}]{Alwall_2014}%
  \BibitemOpen
  \bibfield  {author} {\bibinfo {author} {\bibfnamefont {J.}~\bibnamefont
  {Alwall}}, \bibinfo {author} {\bibfnamefont {R.}~\bibnamefont {Frederix}},
  \bibinfo {author} {\bibfnamefont {S.}~\bibnamefont {Frixione}}, \bibinfo
  {author} {\bibfnamefont {V.}~\bibnamefont {Hirschi}}, \bibinfo {author}
  {\bibfnamefont {F.}~\bibnamefont {Maltoni}}, \bibinfo {author} {\bibfnamefont
  {O.}~\bibnamefont {Mattelaer}}, \bibinfo {author} {\bibfnamefont {H.-S.}\
  \bibnamefont {Shao}}, \bibinfo {author} {\bibfnamefont {T.}~\bibnamefont
  {Stelzer}}, \bibinfo {author} {\bibfnamefont {P.}~\bibnamefont {Torrielli}},\
  and\ \bibinfo {author} {\bibfnamefont {M.}~\bibnamefont {Zaro}},\ }\bibfield
  {title} {\bibinfo {title} {The automated computation of tree-level and
  next-to-leading order differential cross sections, and their matching to
  parton shower simulations},\ }\bibfield  {journal} {\bibinfo  {journal}
  {Journal of High Energy Physics}\ }\textbf {\bibinfo {volume} {2014}},\ \href
  {https://doi.org/10.1007/jhep07(2014)079} {10.1007/jhep07(2014)079} (\bibinfo
  {year} {2014}),\ \Eprint {https://arxiv.org/abs/1405.0301} {arXiv:1405.0301
  [hep-ph]} \BibitemShut {NoStop}%
\bibitem [{\citenamefont {Cacciari}\ \emph {et~al.}(2012)\citenamefont
  {Cacciari}, \citenamefont {Frixione}, \citenamefont {Houdeau}, \citenamefont
  {Mangano}, \citenamefont {Nason},\ and\ \citenamefont
  {Ridolfi}}]{Cacciari_2012}%
  \BibitemOpen
  \bibfield  {author} {\bibinfo {author} {\bibfnamefont {M.}~\bibnamefont
  {Cacciari}}, \bibinfo {author} {\bibfnamefont {S.}~\bibnamefont {Frixione}},
  \bibinfo {author} {\bibfnamefont {N.}~\bibnamefont {Houdeau}}, \bibinfo
  {author} {\bibfnamefont {M.~L.}\ \bibnamefont {Mangano}}, \bibinfo {author}
  {\bibfnamefont {P.}~\bibnamefont {Nason}},\ and\ \bibinfo {author}
  {\bibfnamefont {G.}~\bibnamefont {Ridolfi}},\ }\bibfield  {title} {\bibinfo
  {title} {Theoretical predictions for charm and bottom production at the
  lhc},\ }\bibfield  {journal} {\bibinfo  {journal} {Journal of High Energy
  Physics}\ }\textbf {\bibinfo {volume} {2012}},\ \href
  {https://doi.org/10.1007/jhep10(2012)137} {10.1007/jhep10(2012)137} (\bibinfo
  {year} {2012}),\ \Eprint {https://arxiv.org/abs/1205.6344} {arXiv:1205.6344
  [hep-ph]} \BibitemShut {NoStop}%
\bibitem [{\citenamefont {Cacciari}()}]{FONLL}%
  \BibitemOpen
  \bibfield  {author} {\bibinfo {author} {\bibfnamefont {M.}~\bibnamefont
  {Cacciari}},\ }\href@noop {} {\bibinfo {title} {Fonll heavy quark
  production}},\ \Eprint
  {https://arxiv.org/abs/http://www.lpthe.jussieu.fr/~cacciari/fonll/fonllform.html}
  {http://www.lpthe.jussieu.fr/~cacciari/fonll/fonllform.html} \BibitemShut
  {NoStop}%
\bibitem [{TWI()}]{TWIKI}%
  \BibitemOpen
  \href@noop {} {\bibinfo {title} {Cernyellowreportpageat1314tev2014}},\
  \Eprint
  {https://arxiv.org/abs/https://twiki.cern.ch/twiki/bin/view/LHCPhysics/}
  {https://twiki.cern.ch/twiki/bin/view/LHCPhysics/} \BibitemShut {NoStop}%
\bibitem [{\citenamefont {Evans}(2018)}]{Evans_2018}%
  \BibitemOpen
  \bibfield  {author} {\bibinfo {author} {\bibfnamefont {J.~A.}\ \bibnamefont
  {Evans}},\ }\bibfield  {title} {\bibinfo {title} {Detecting hidden particles
  with mathusla},\ }\bibfield  {journal} {\bibinfo  {journal} {Physical Review
  D}\ }\textbf {\bibinfo {volume} {97}},\ \href
  {https://doi.org/10.1103/physrevd.97.055046} {10.1103/physrevd.97.055046}
  (\bibinfo {year} {2018}),\ \Eprint {https://arxiv.org/abs/1708.08503}
  {arXiv:1708.08503 [hep-ph]} \BibitemShut {NoStop}%
\end{thebibliography}%

\newpage
\onecolumngrid

\appendix

\setcounter{figure}{0}
\counterwithin{figure}{section}

\section{Decay Probability Maps}

LLP decay probability maps are shown here, along with geometric optimization parameters. They
show values of $P_{map,i}$
 for 5m $ \times $ 5m squares, as described in section II.C. All distances are in meters, and all areas are in meters squared. They appear here in the order provided by Table III. 
\begin{figure}[h]
\includegraphics[scale=0.65]{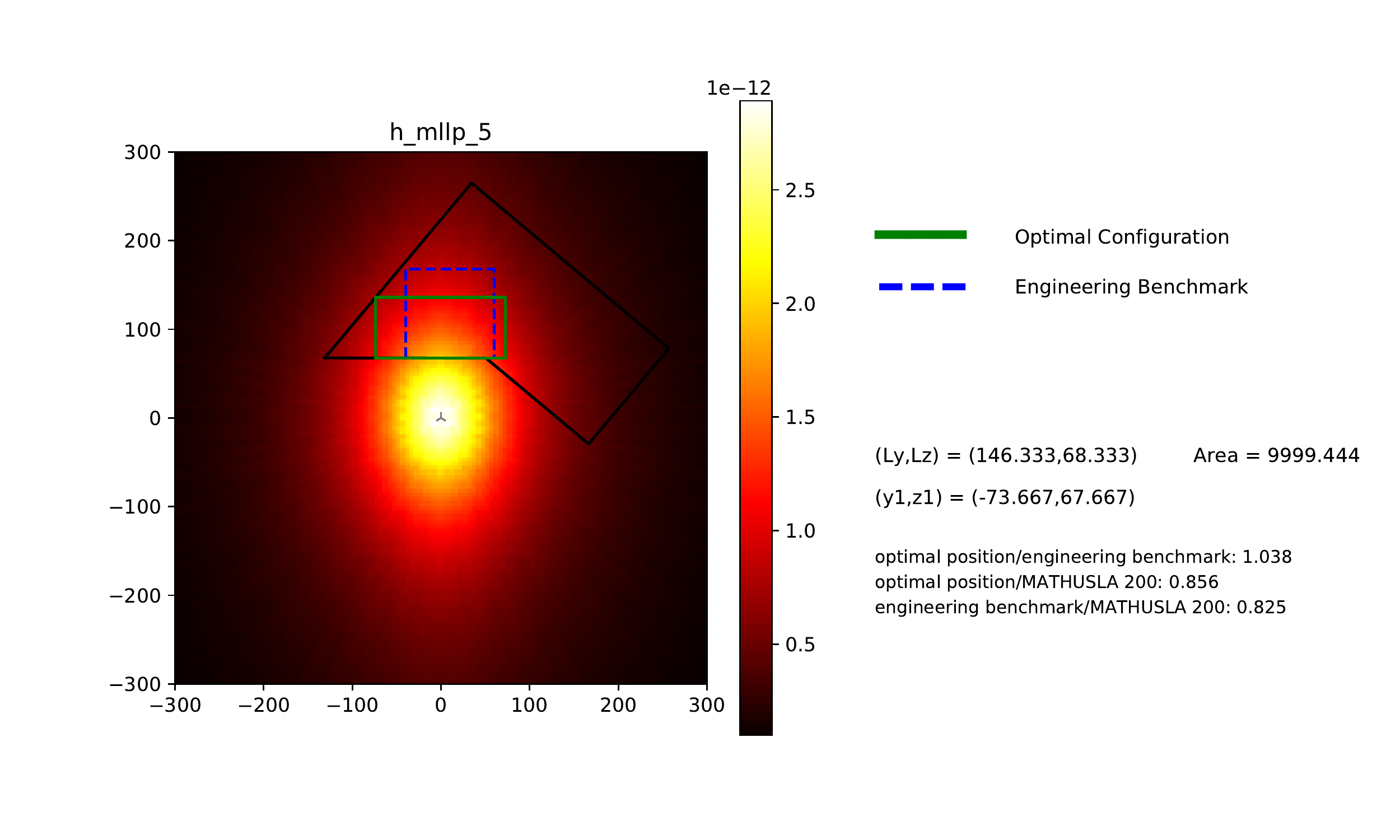}% Here is how to import EPS art
\caption{\label{fig:epsart}  HAHM, $m_{LLP} = 5$ GeV }
\end{figure}
\begin{figure}[h]
\includegraphics[scale=0.65]{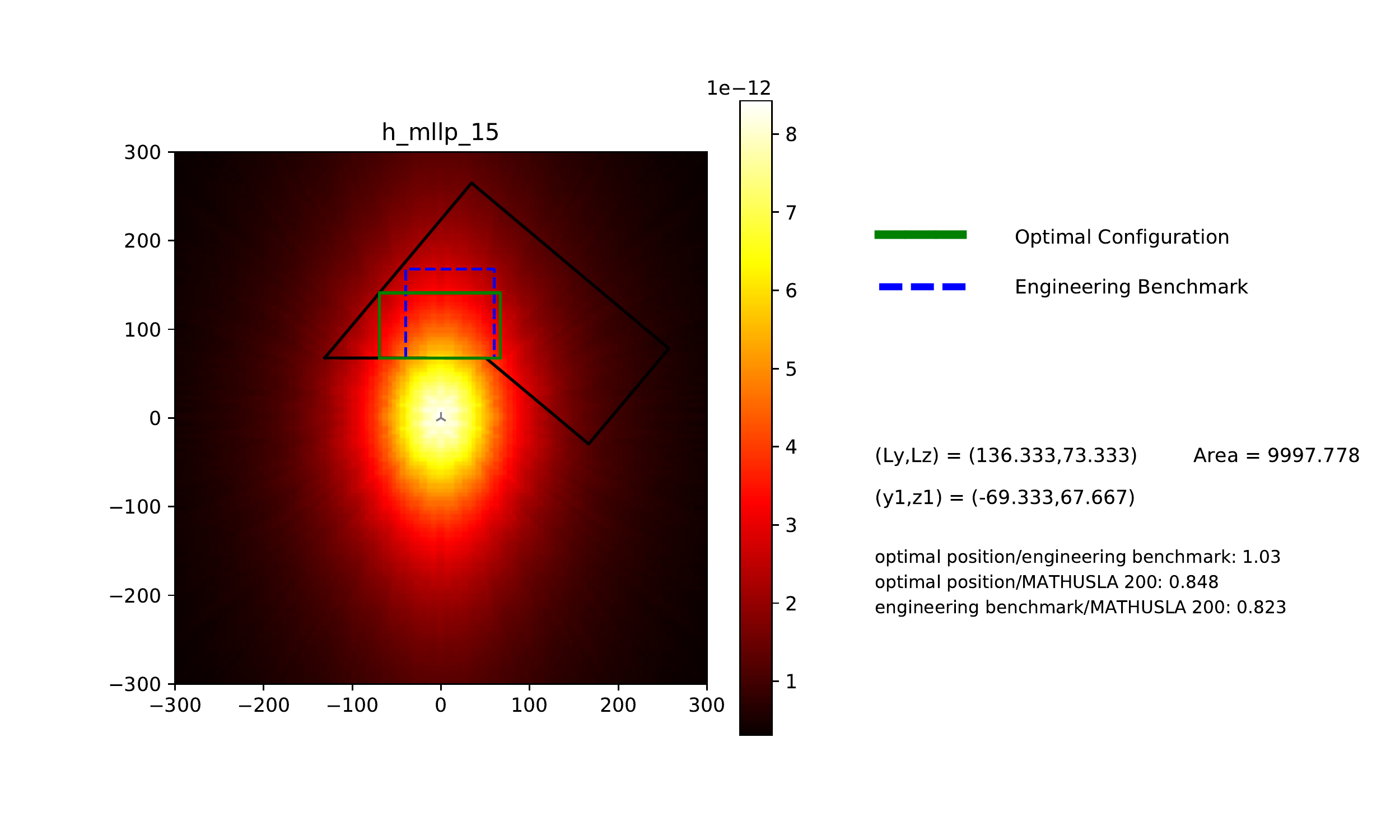}% Here is how to import EPS art
\caption{\label{fig:epsart}  HAHM, $m_{LLP} = 15$ GeV }
\end{figure}
\begin{figure}[h]
\includegraphics[scale=0.65]{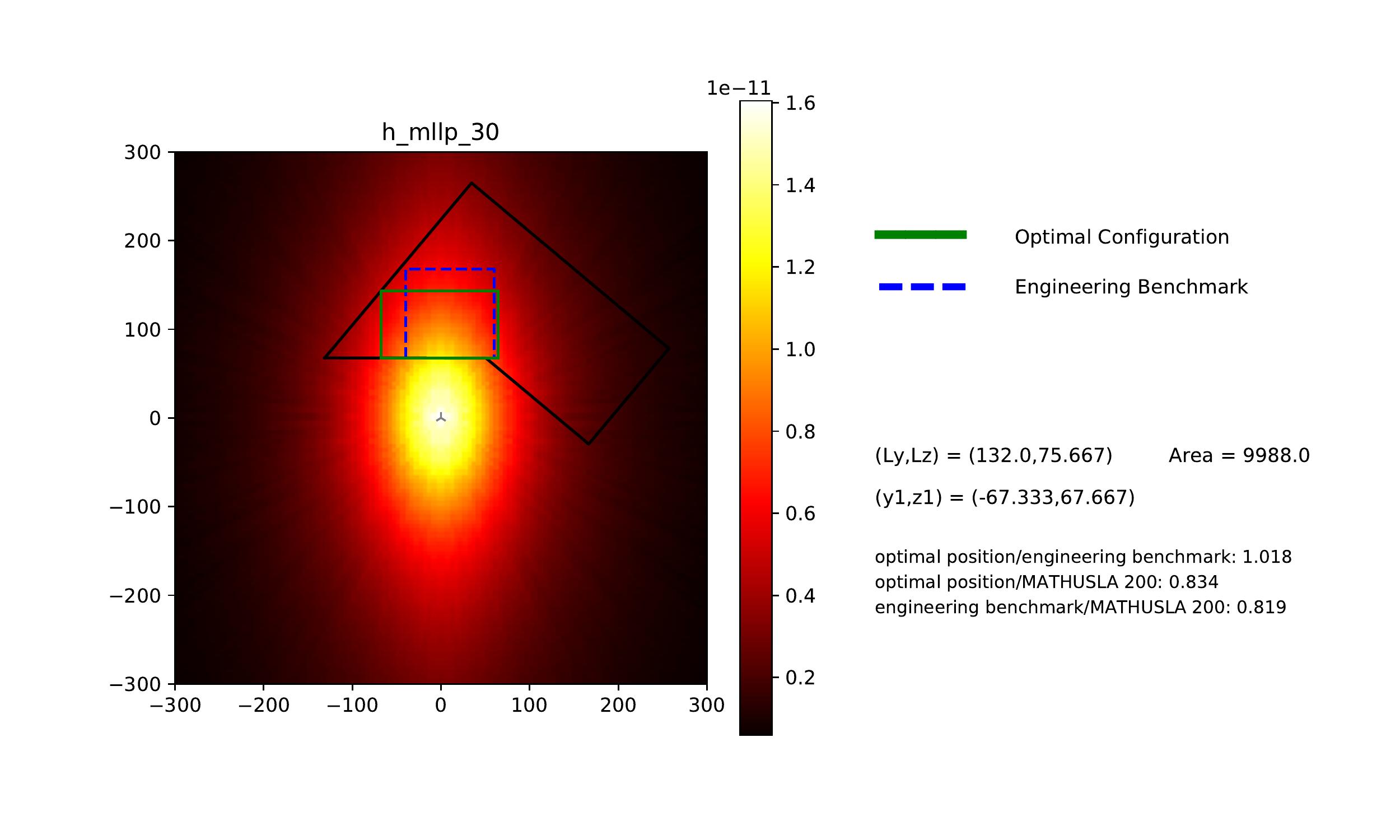}% Here is how to import EPS art
\caption{\label{fig:epsart} HAHM, $m_{LLP} = 30$ GeV  }
\end{figure}
\begin{figure}[h]
\includegraphics[scale=0.65]{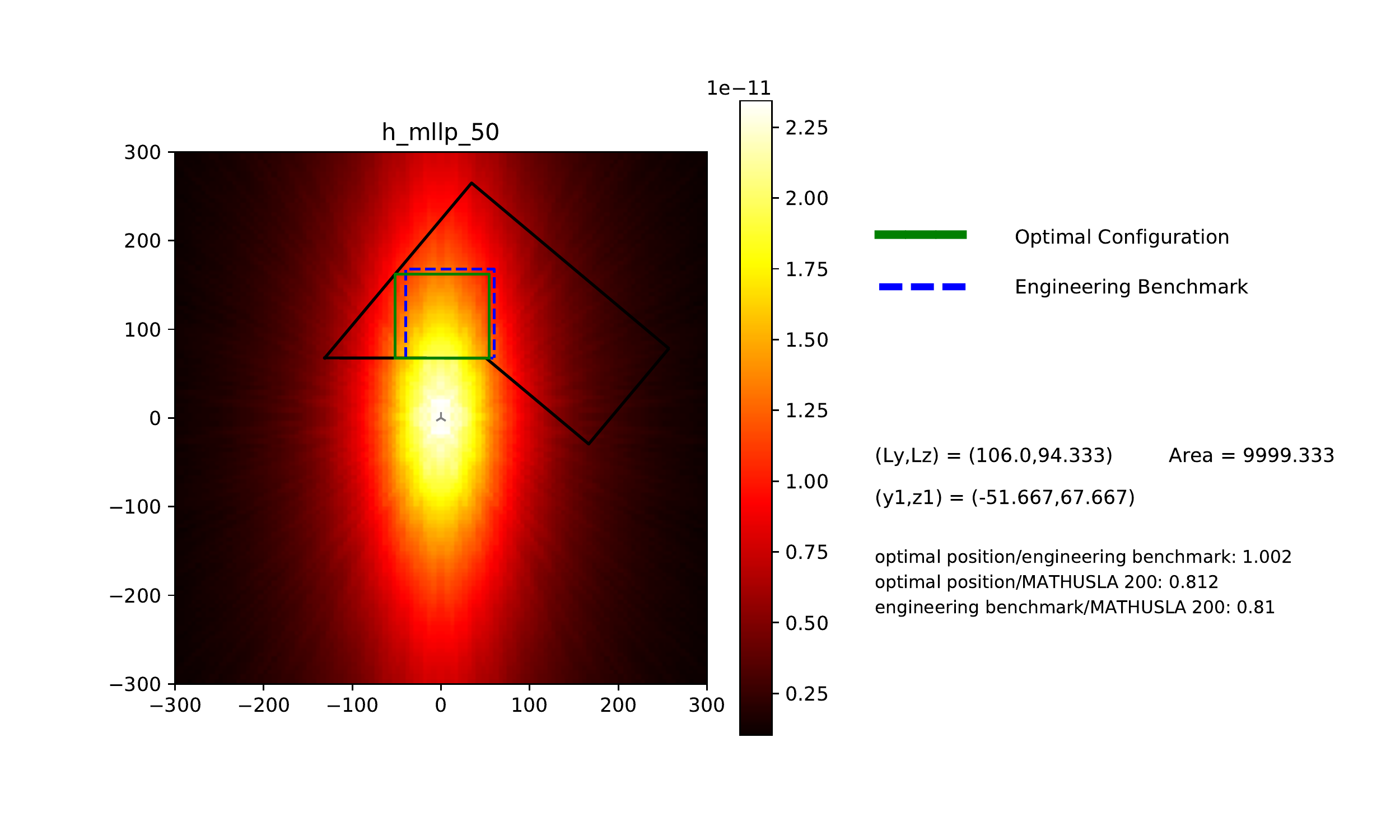}% Here is how to import EPS art
\caption{\label{fig:epsart} HAHM, $m_{LLP} = 50$ GeV }
\end{figure}
\begin{figure}[h]
\includegraphics[scale=0.65]{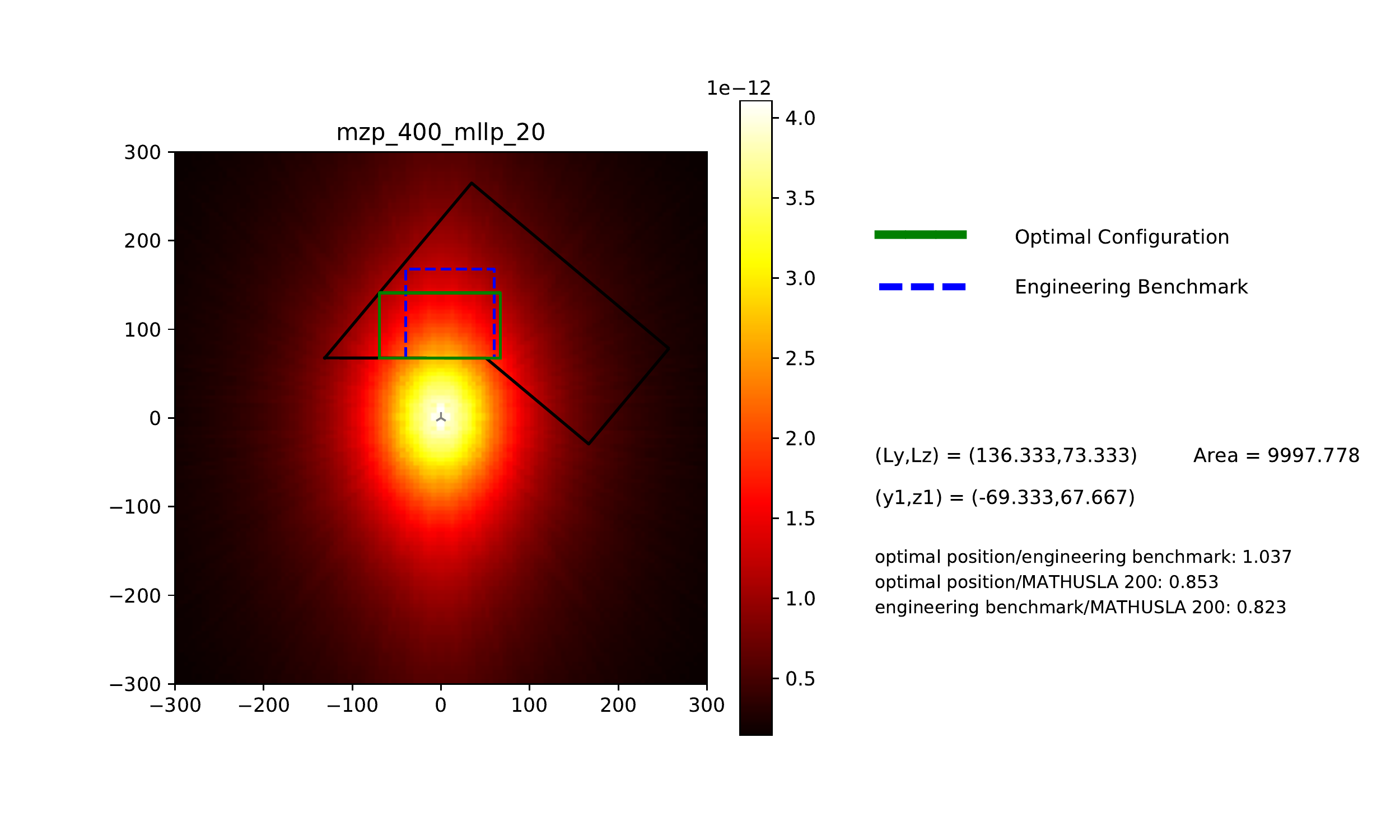}% Here is how to import EPS art
\caption{\label{fig:epsart}  ZP, $m_{ZP} = 0.4 \quad TeV$, $m_{LLP} = 20$ GeV }
\end{figure}
\begin{figure}[h]
\includegraphics[scale=0.65]{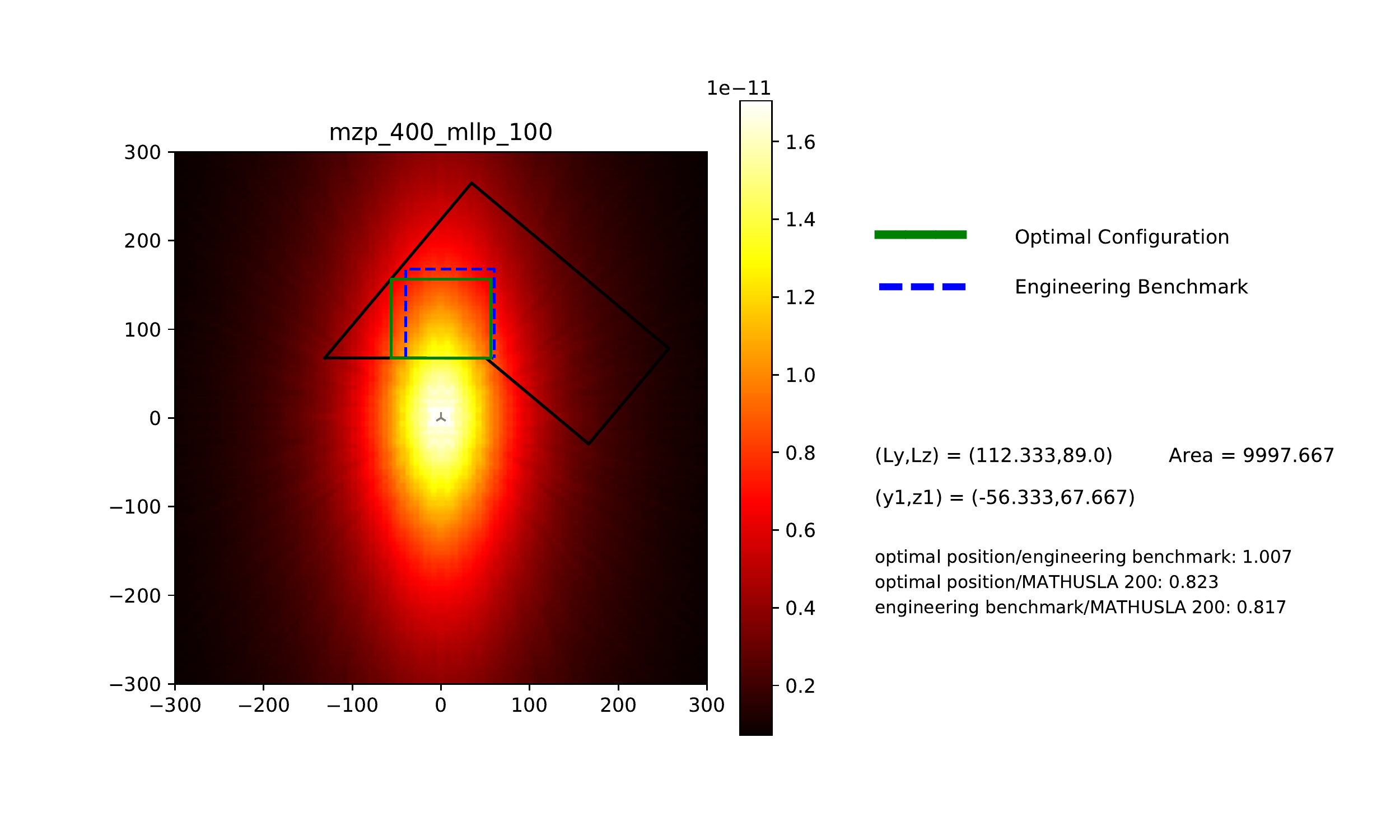}% Here is how to import EPS art
\caption{\label{fig:epsart}  ZP, $m_{ZP} = 0.4 \quad TeV$, $m_{LLP} = 100$ GeV }
\end{figure}
\begin{figure}[h]
\includegraphics[scale=0.65]{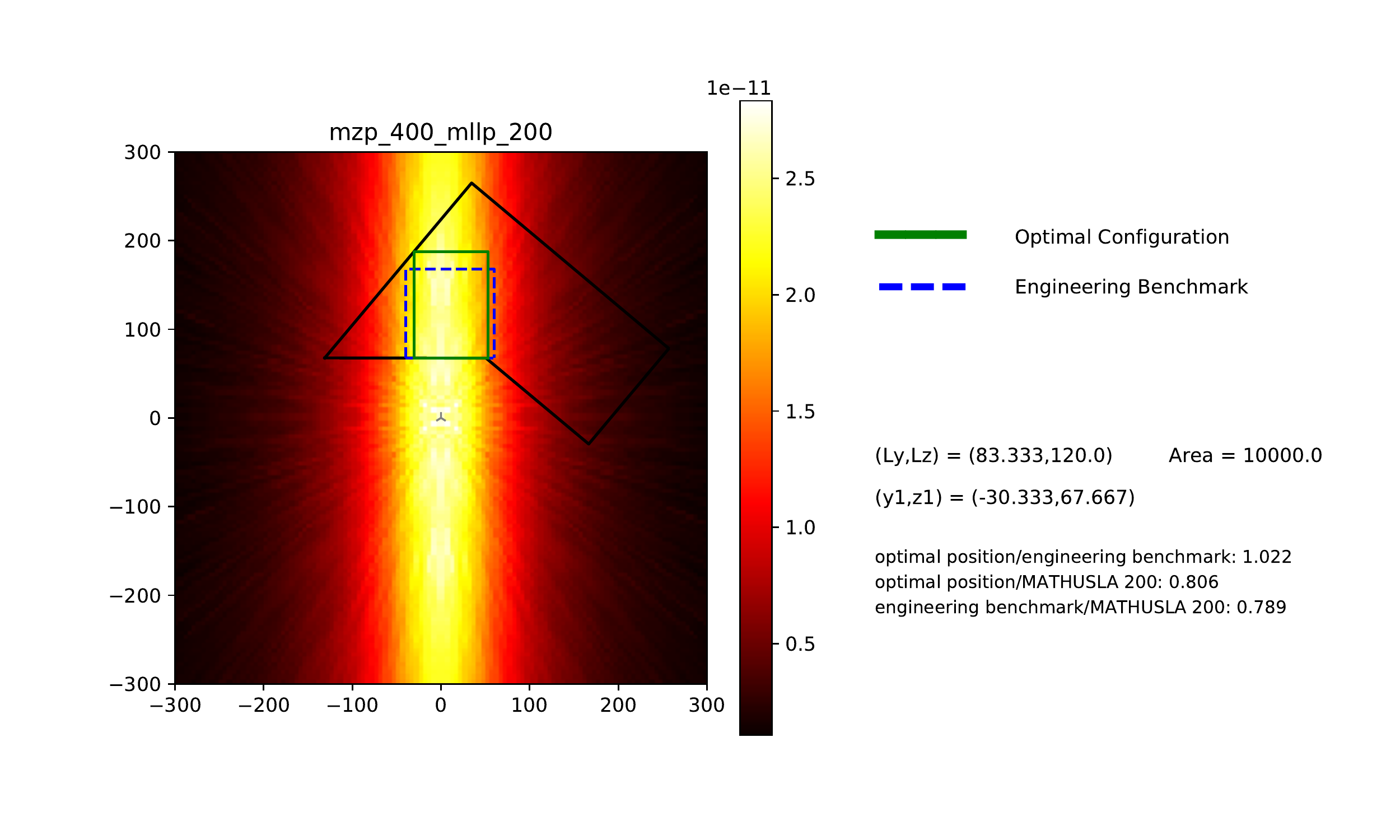}% Here is how to import EPS art
\caption{\label{fig:epsart}  ZP, $m_{ZP} = 0.4 \quad TeV$, $m_{LLP} = 200$ GeV }
\end{figure}
\begin{figure}[h]
\includegraphics[scale=0.65]{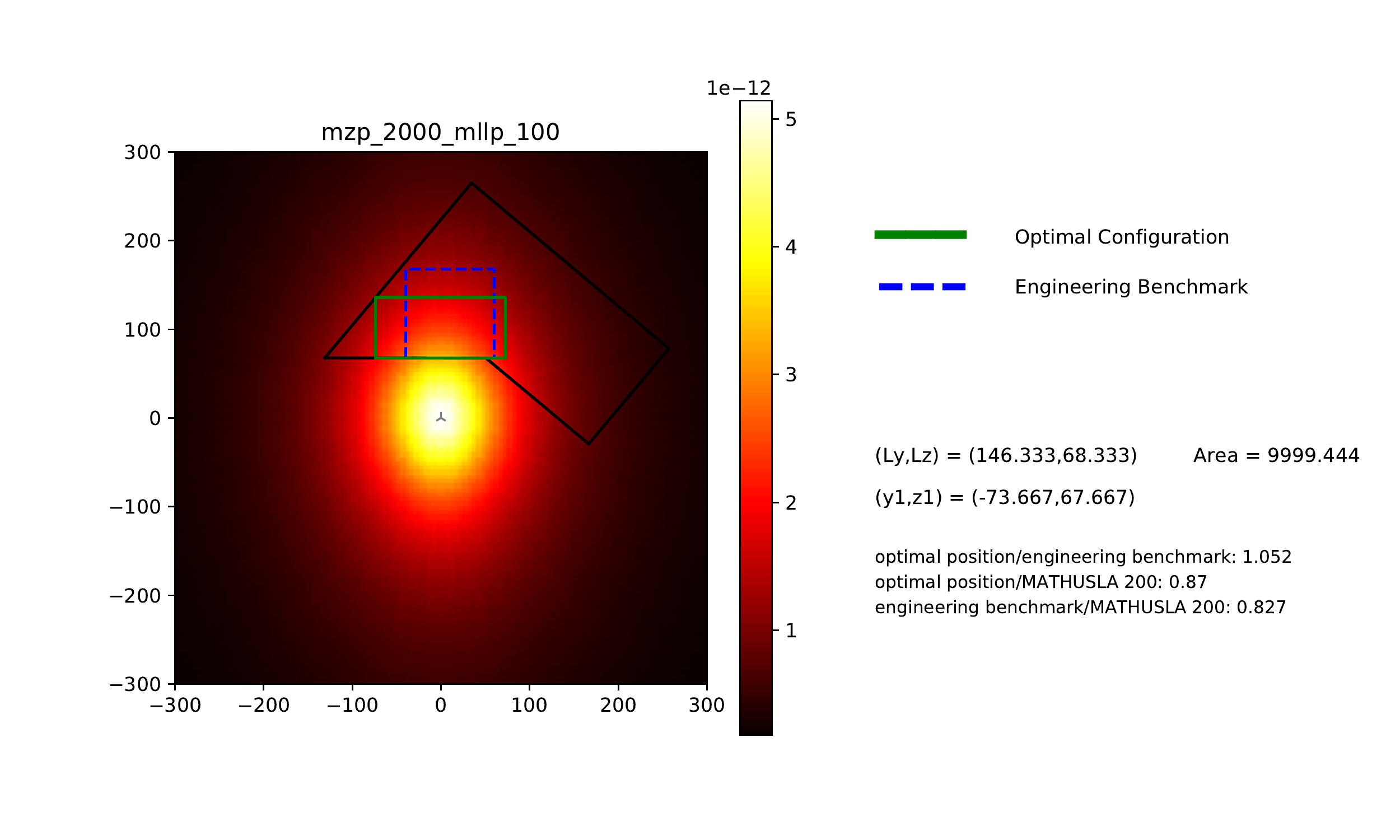}% Here is how to import EPS art
\caption{\label{fig:epsart}  ZP, $m_{ZP} = 2 \quad TeV$, $m_{LLP} = 100$ GeV }
\end{figure}
\begin{figure}[h]
\includegraphics[scale=0.65]{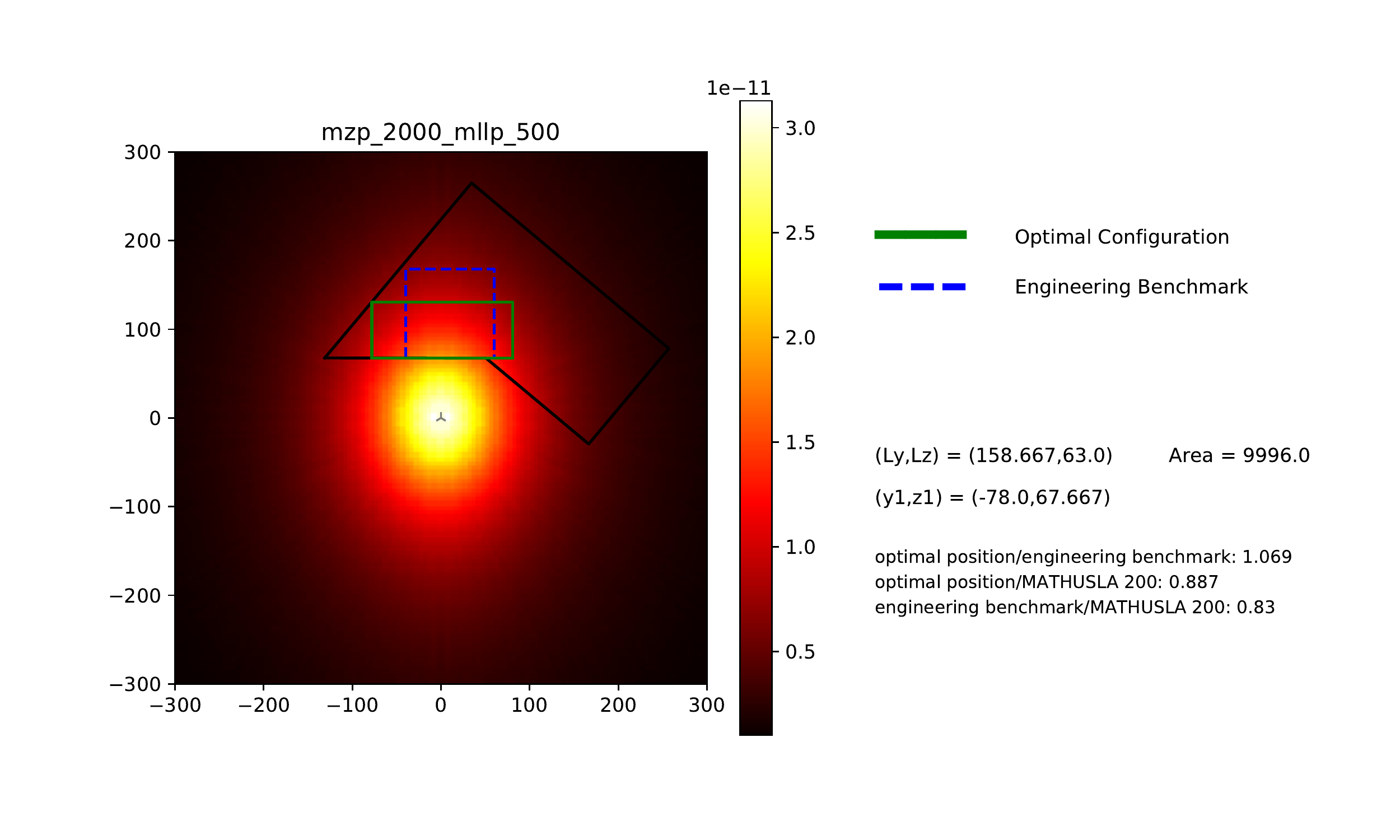}% Here is how to import EPS art
\caption{\label{fig:epsart}  ZP, $m_{ZP} = 2 \quad TeV$, $m_{LLP} = 500$ GeV }
\end{figure}
\begin{figure}[h]
\includegraphics[scale=0.65]{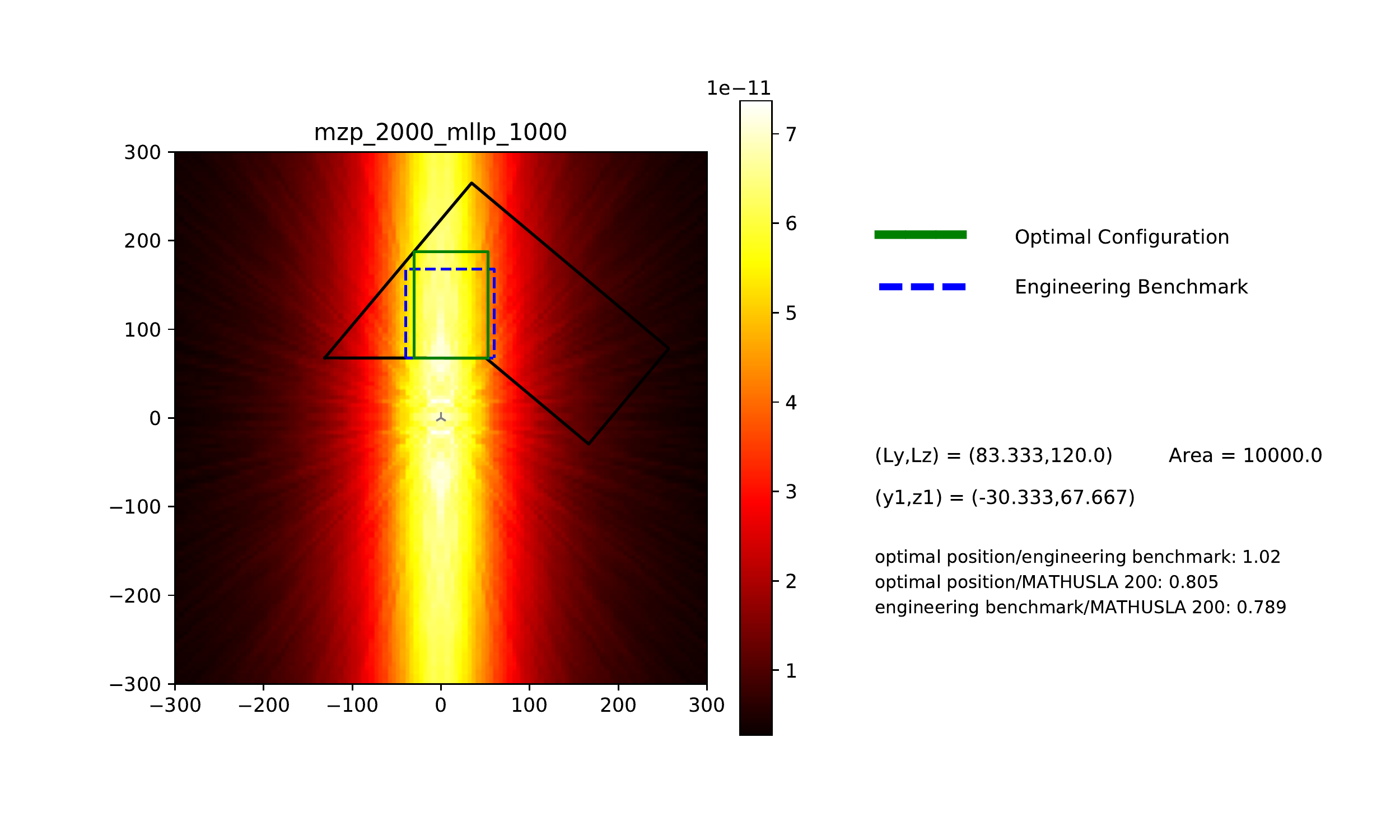}% Here is how to import EPS art
\caption{\label{fig:epsart}  ZP, $m_{ZP} = 2 \quad TeV$, $m_{LLP} = 1000$ GeV }
\end{figure}
\begin{figure}[h]
\includegraphics[scale=0.65]{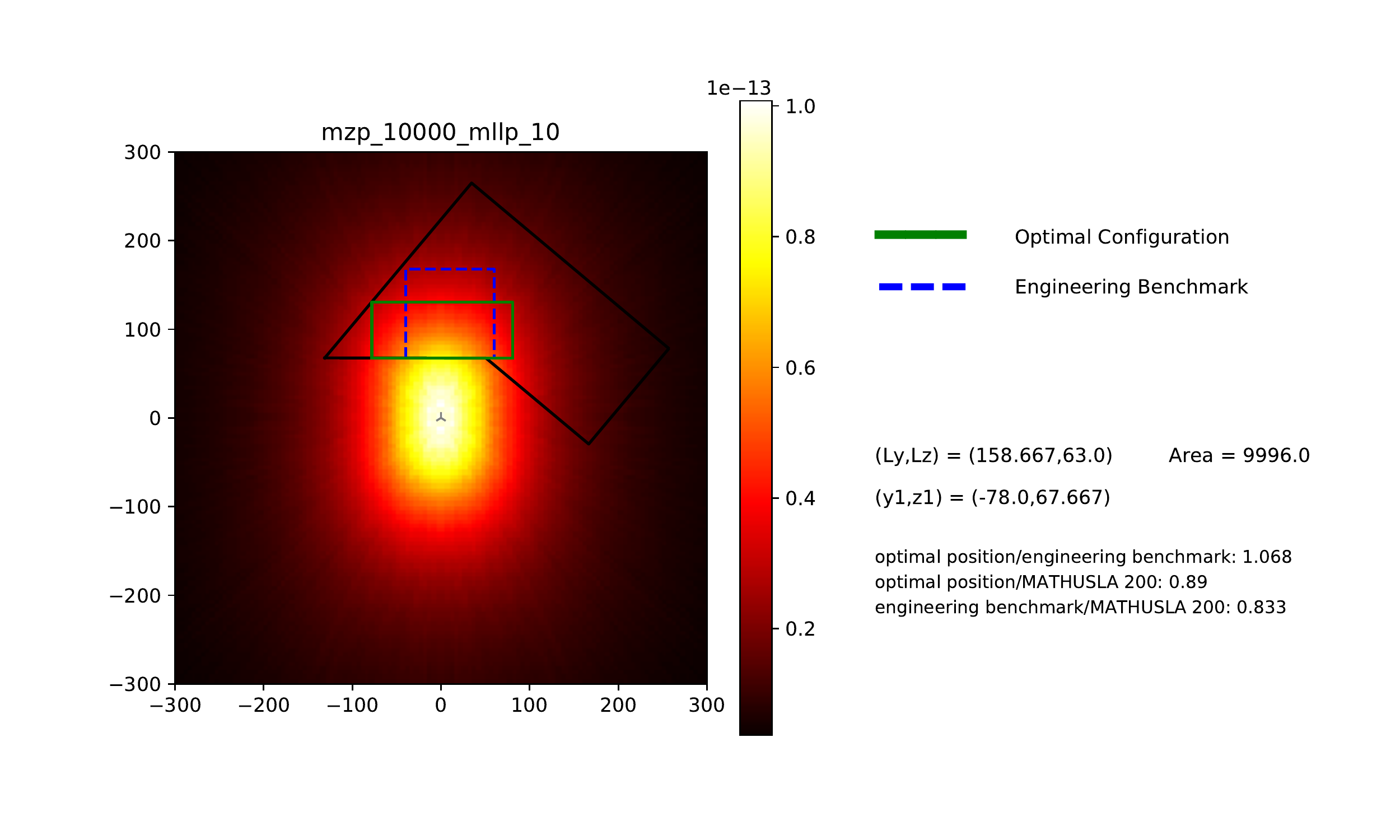}% Here is how to import EPS art
\caption{\label{fig:epsart}  ZP, $m_{ZP} = 10 \quad TeV$, $m_{LLP} = 10$ GeV }
\end{figure}
\begin{figure}[h]
\includegraphics[scale=0.65]{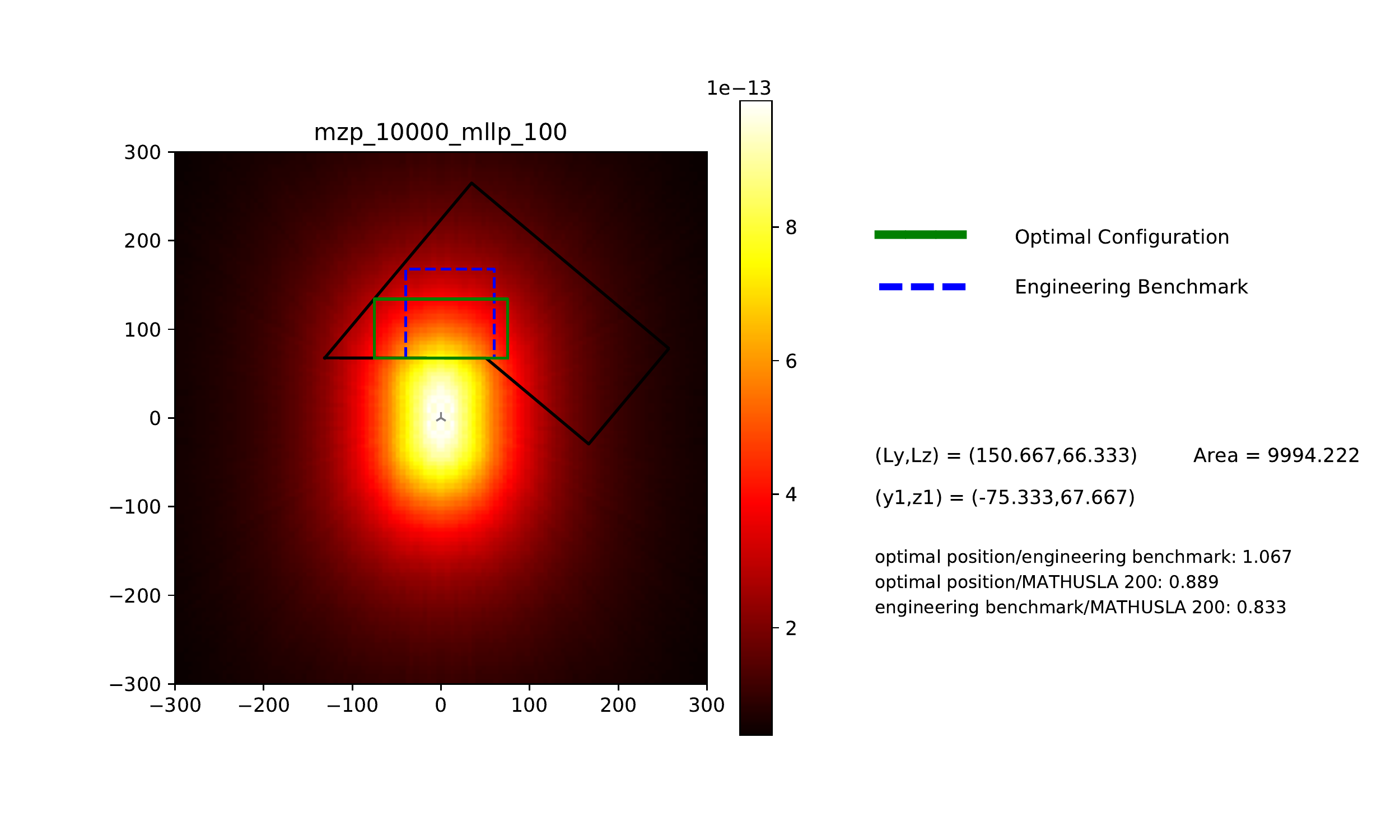}% Here is how to import EPS art
\caption{\label{fig:epsart} ZP, $m_{ZP} = 10 \quad TeV$, $m_{LLP} = 100$ GeV  }
\end{figure}
\begin{figure}[h]
\includegraphics[scale=0.65]{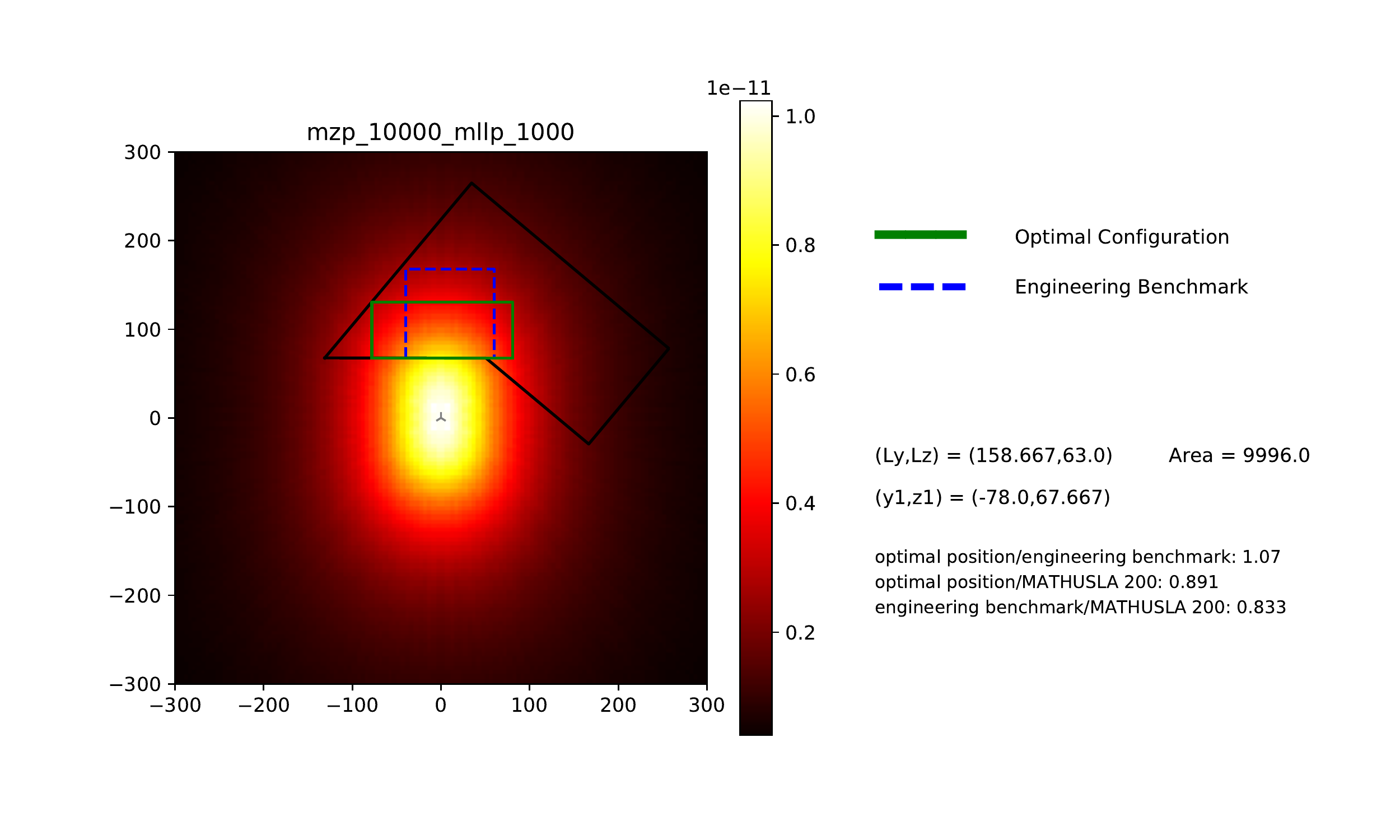}% Here is how to import EPS art
\caption{\label{fig:epsart}  ZP, $m_{ZP} = 10 \quad TeV$, $m_{LLP} = 1000$ GeV }
\end{figure}
\begin{figure}[h]
\includegraphics[scale=0.65]{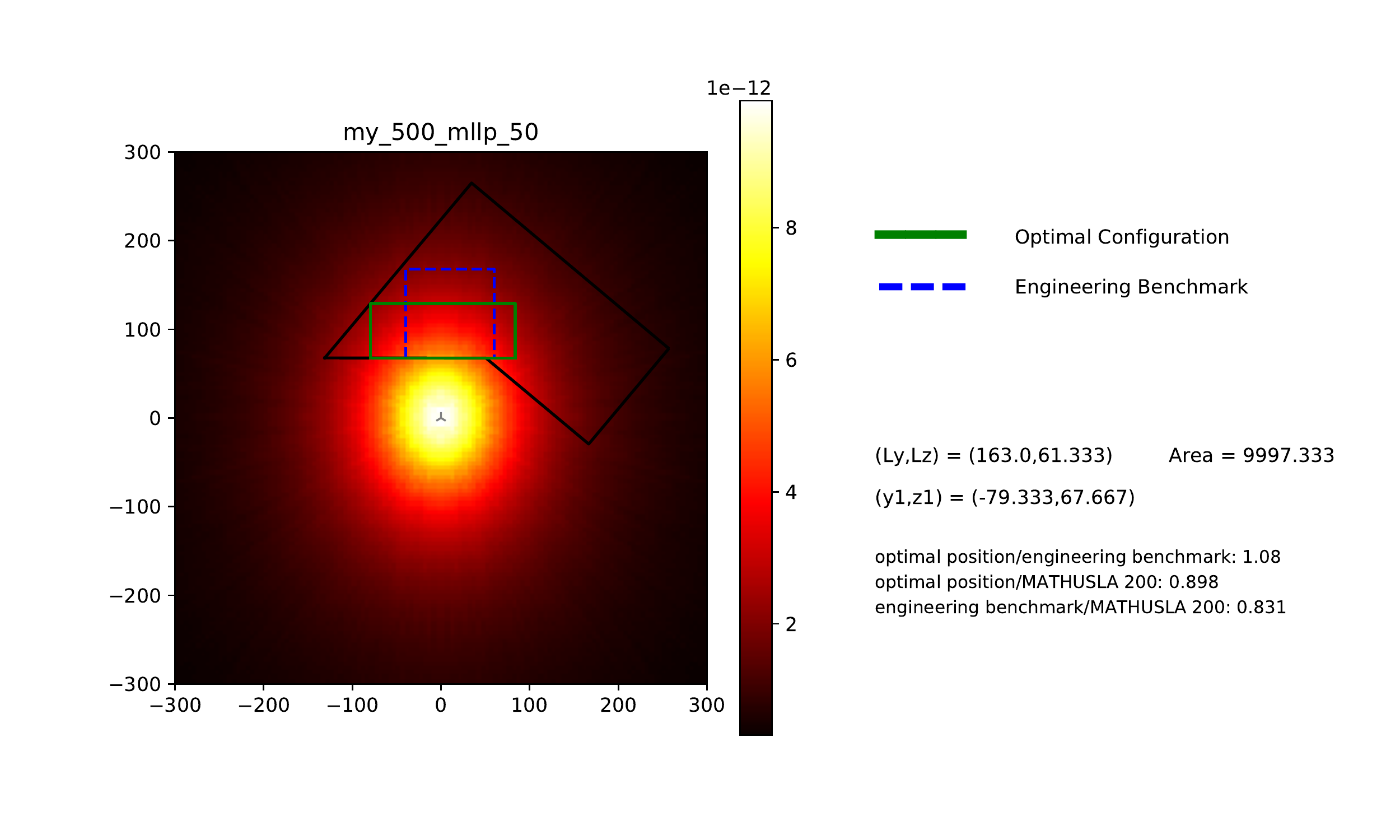}% Here is how to import EPS art
\caption{\label{fig:epsart}  RPVMSSM, $m_y = 0.5 \quad TeV$, $m_{LLP} = 50$ GeV }
\end{figure}
\begin{figure}[h]
\includegraphics[scale=0.65]{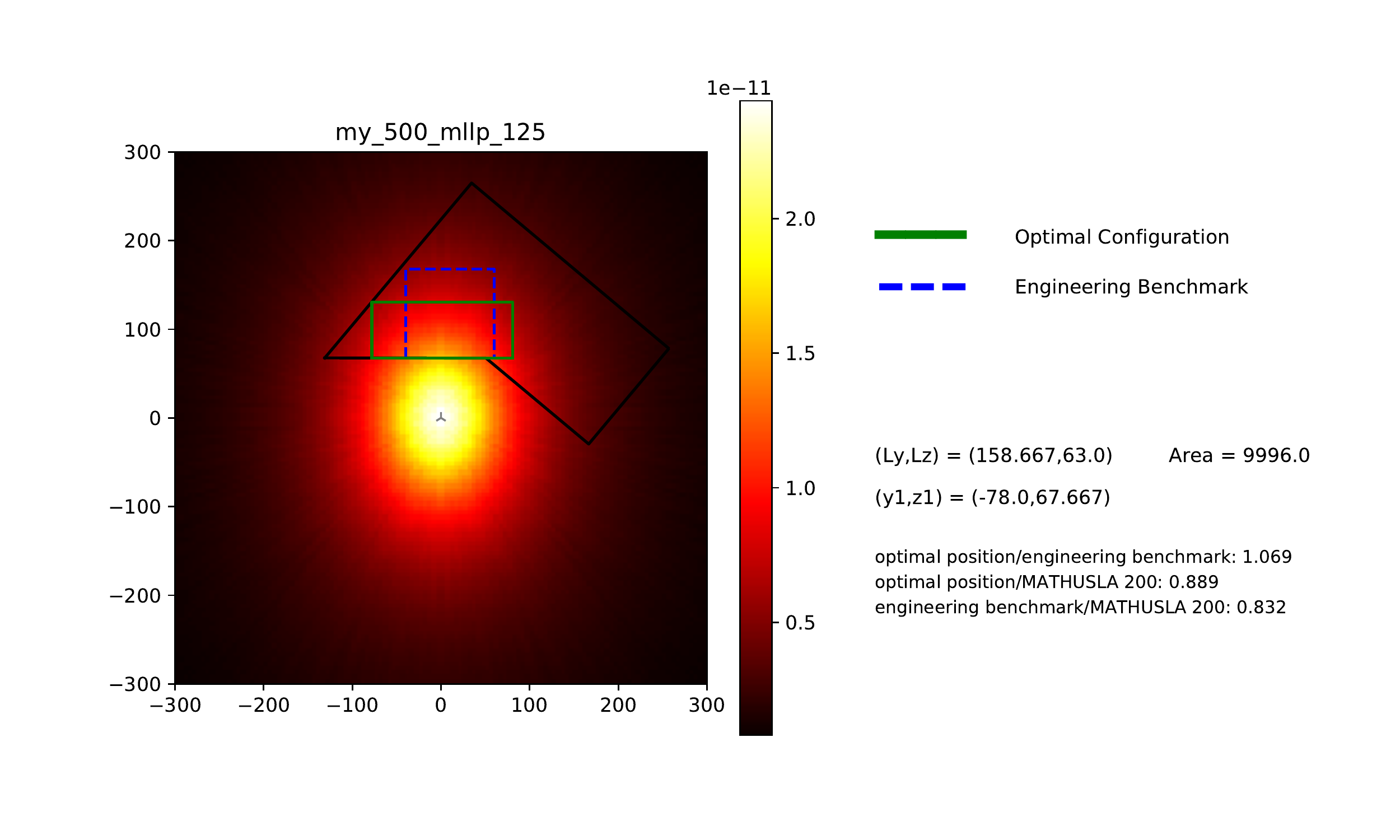}% Here is how to import EPS art
\caption{\label{fig:epsart}  RPVMSSM, $m_y = 0.5 \quad TeV$, $m_{LLP} = 125$ GeV }
\end{figure}
\begin{figure}[h]
\includegraphics[scale=0.65]{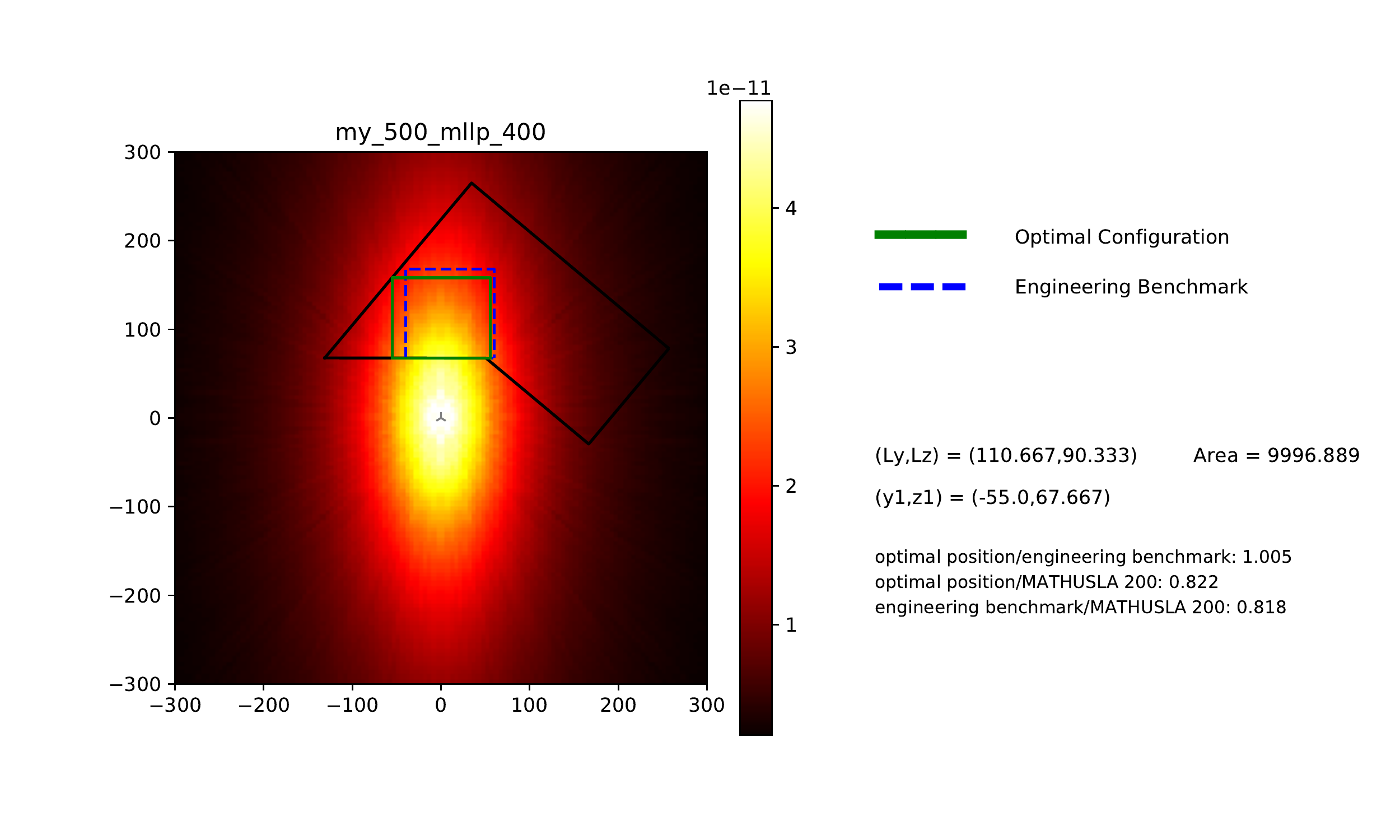}% Here is how to import EPS art
\caption{\label{fig:epsart}  RPVMSSM, $m_y = 0.5 \quad TeV$, $m_{LLP} = 400$ GeV }
\end{figure}
\begin{figure}[h]
\includegraphics[scale=0.65]{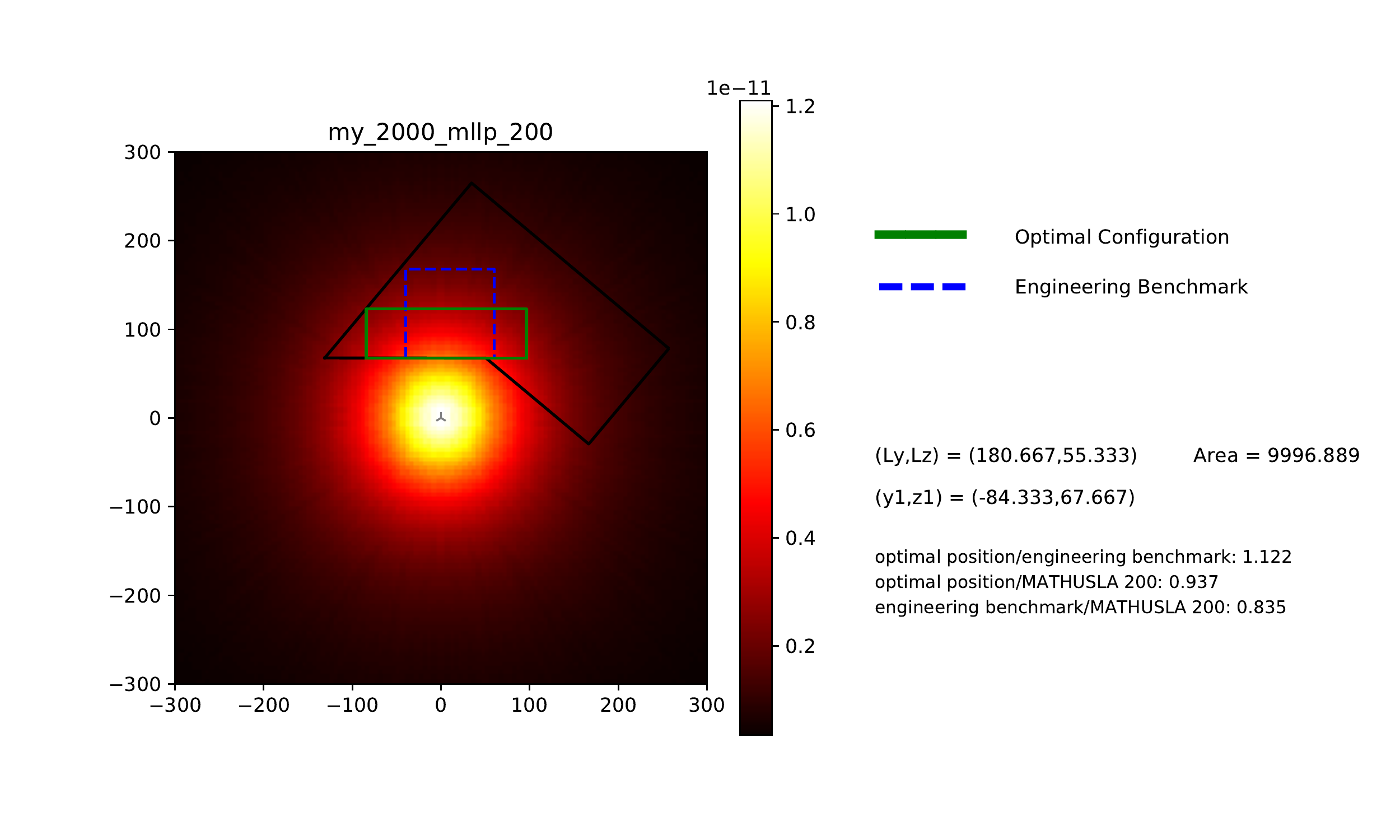}% Here is how to import EPS art
\caption{\label{fig:epsart}  RPVMSSM, $m_y = 2 \quad TeV$, $m_{LLP} = 200$ GeV }
\end{figure}
\begin{figure}[h]
\includegraphics[scale=0.65]{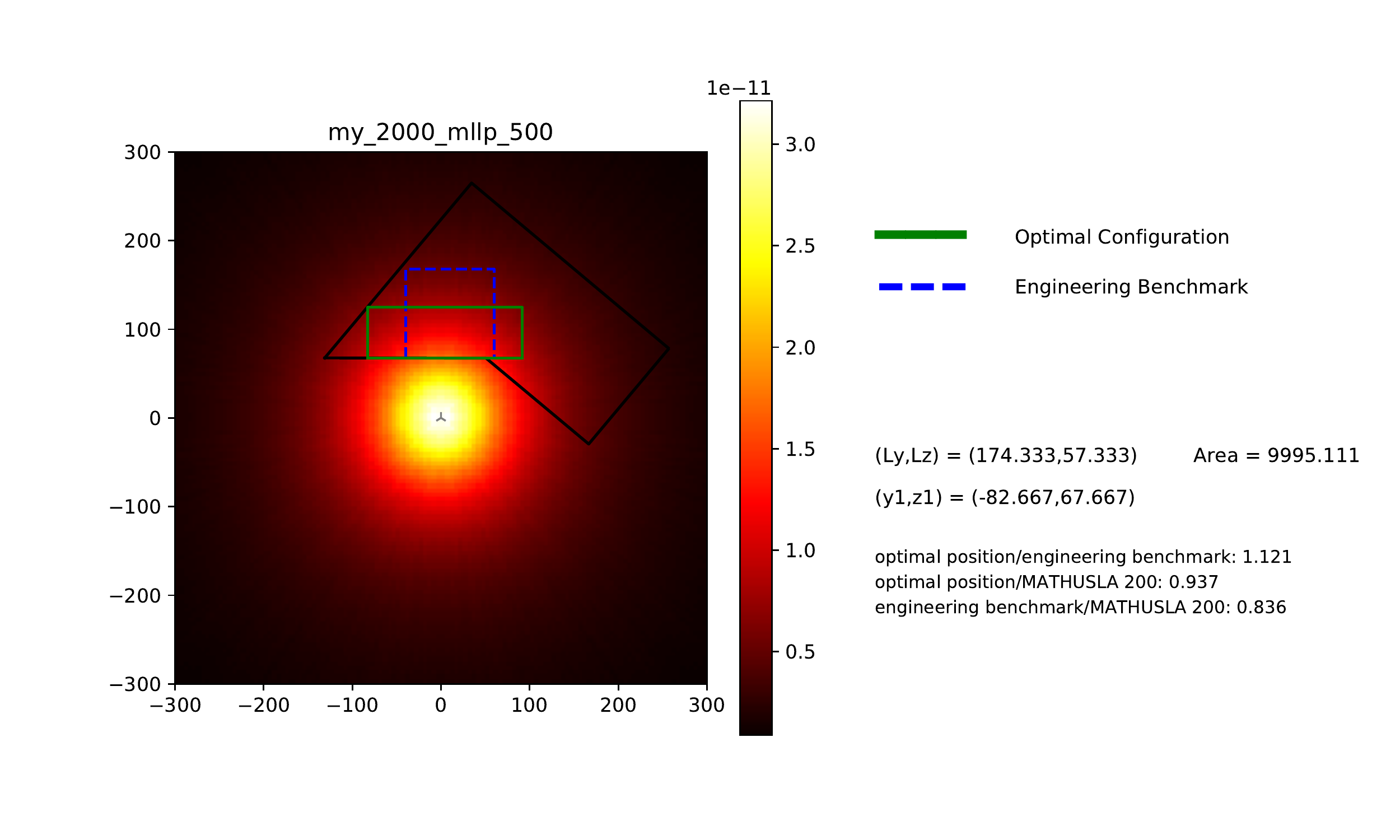}% Here is how to import EPS art
\caption{\label{fig:epsart}  RPVMSSM, $m_y = 2 \quad TeV$, $m_{LLP} = 500$ GeV }
\end{figure}
\begin{figure}[h]
\includegraphics[scale=0.65]{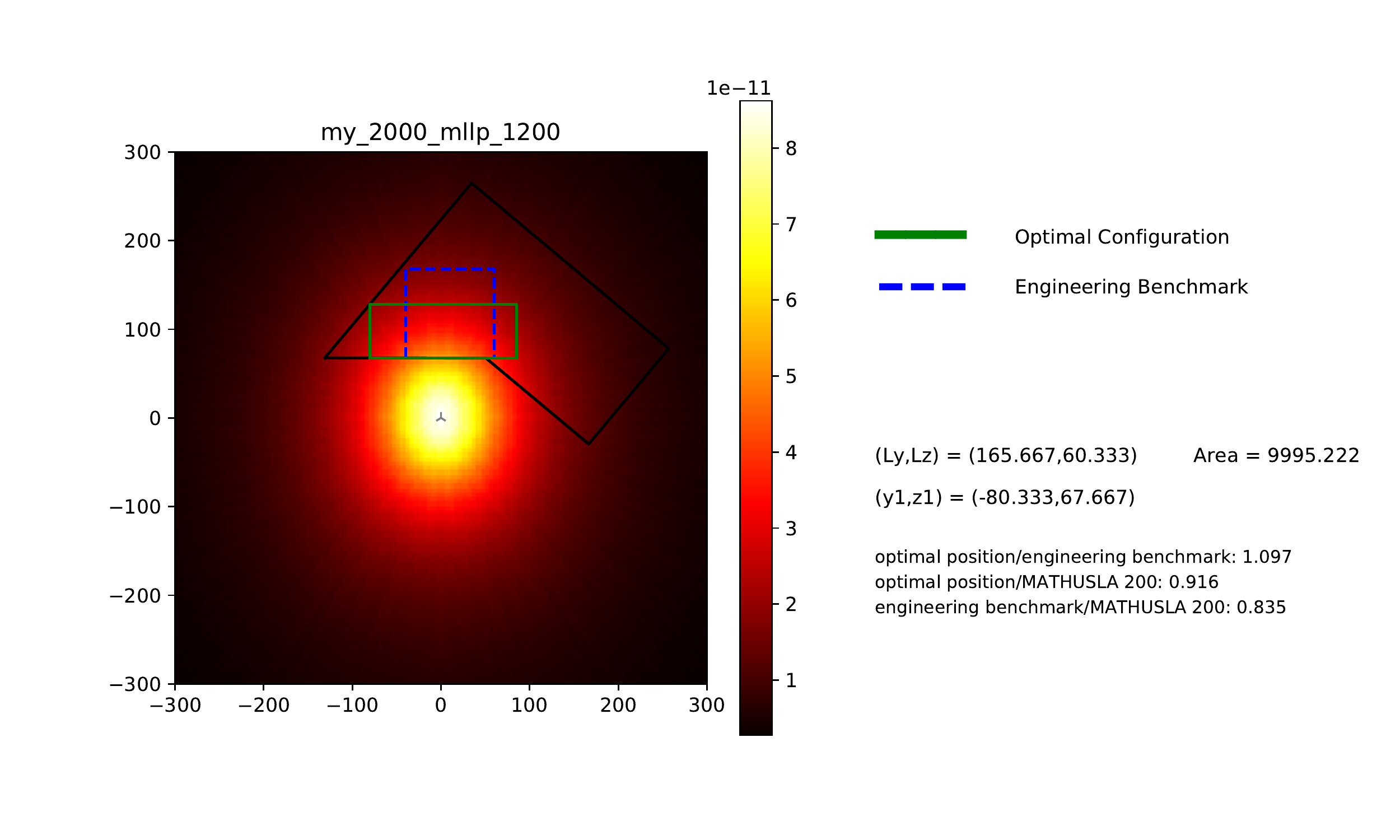}% Here is how to import EPS art
\caption{\label{fig:epsart}  RPVMSSM, $m_y = 2 \quad TeV$, $m_{LLP} = 1200$ GeV }
\end{figure}
\begin{figure}[h]
\includegraphics[scale=0.65]{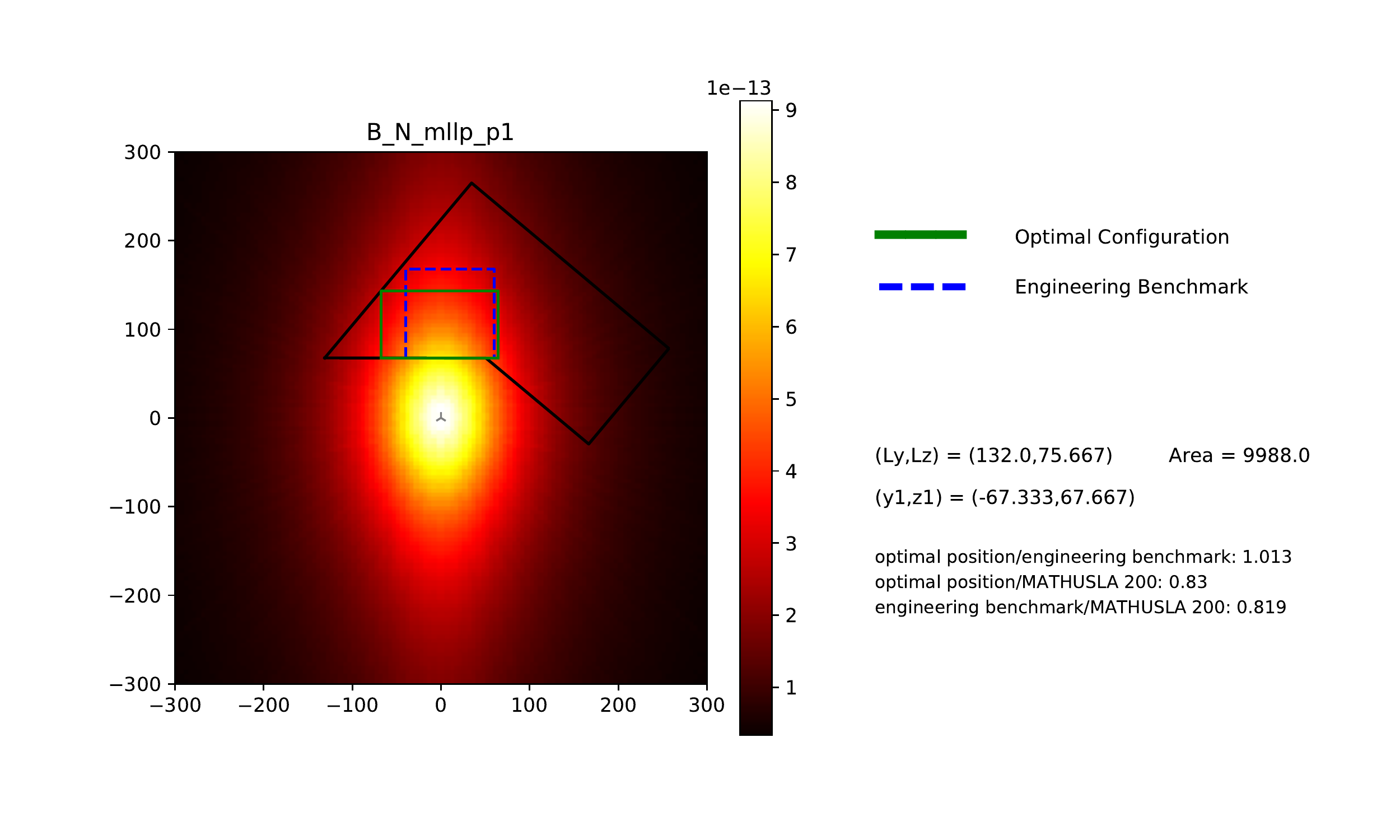}% Here is how to import EPS art
\caption{\label{fig:epsart}  p p $\rightarrow$ B $\rightarrow$ RHN, $m_{LLP} = 0.1$ GeV }
\end{figure}
\begin{figure}[h]
\includegraphics[scale=0.65]{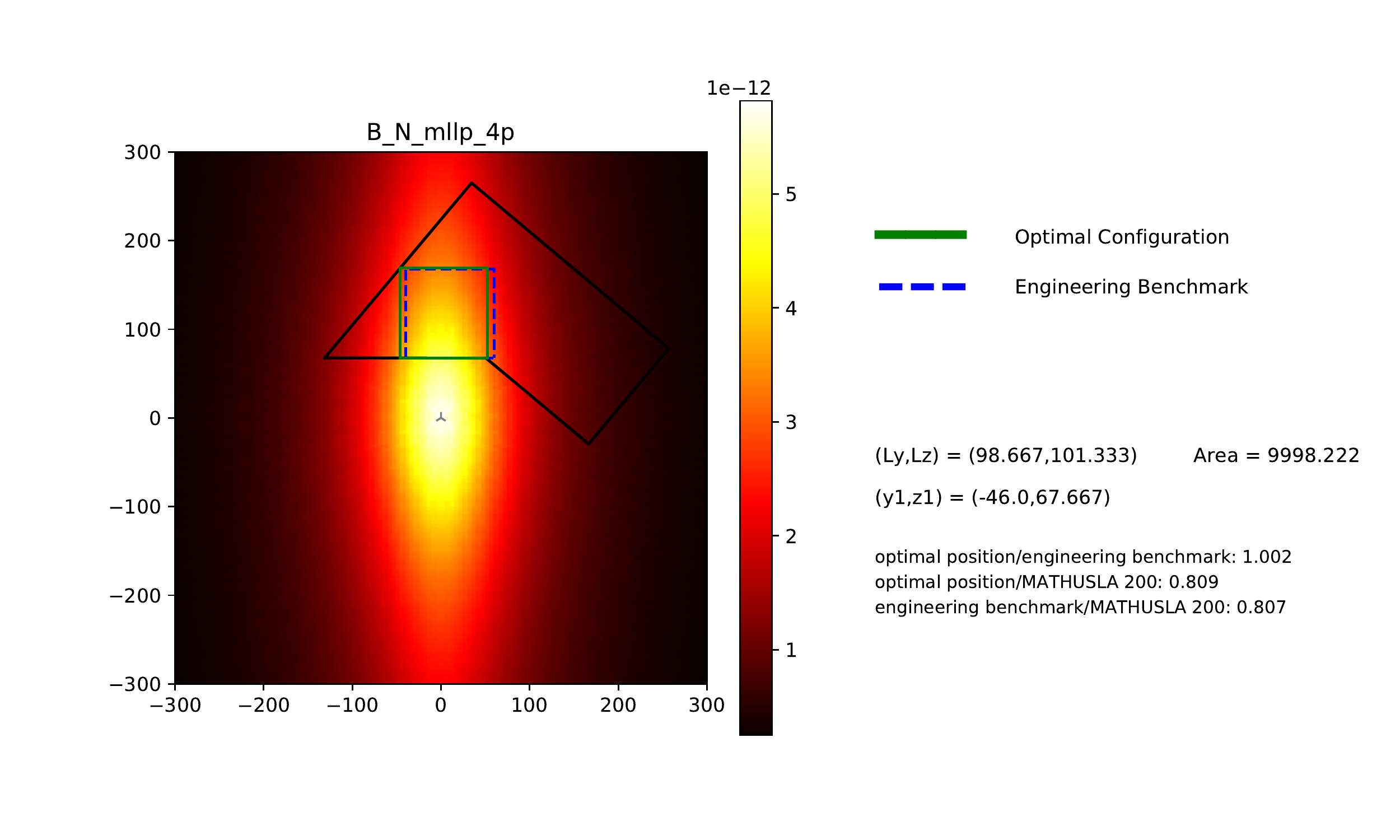}% Here is how to import EPS art
\caption{\label{fig:epsart}  p p $\rightarrow$ B $\rightarrow$ RHN, $m_{LLP} = 4.0$ GeV }
\end{figure}
\begin{figure}[h]
\includegraphics[scale=0.65]{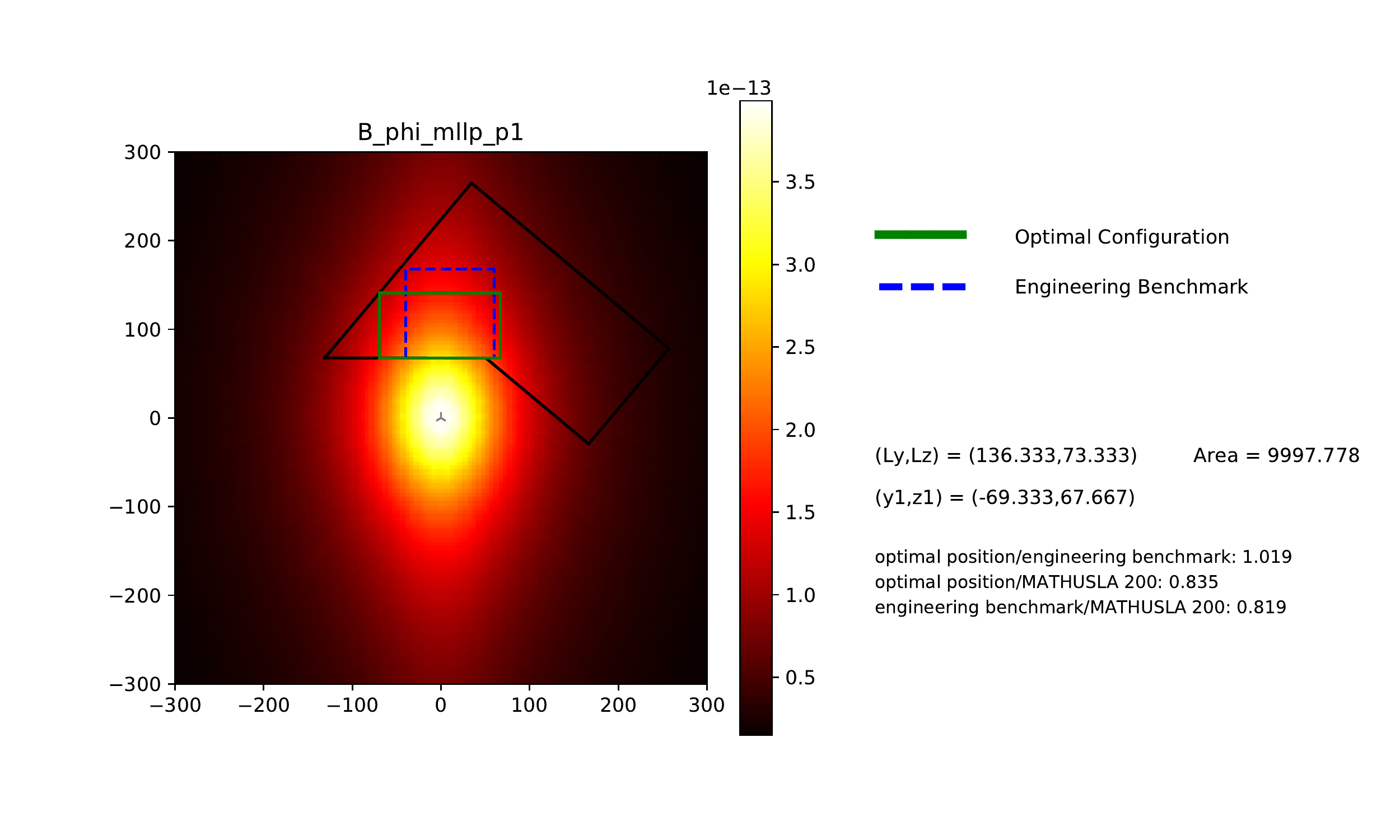}% Here is how to import EPS art
\caption{\label{fig:epsart}  p p $\rightarrow$ B $\rightarrow$ scalar, $m_{LLP} = 0.1$ GeV }
\end{figure}
\begin{figure}[h]
\includegraphics[scale=0.65]{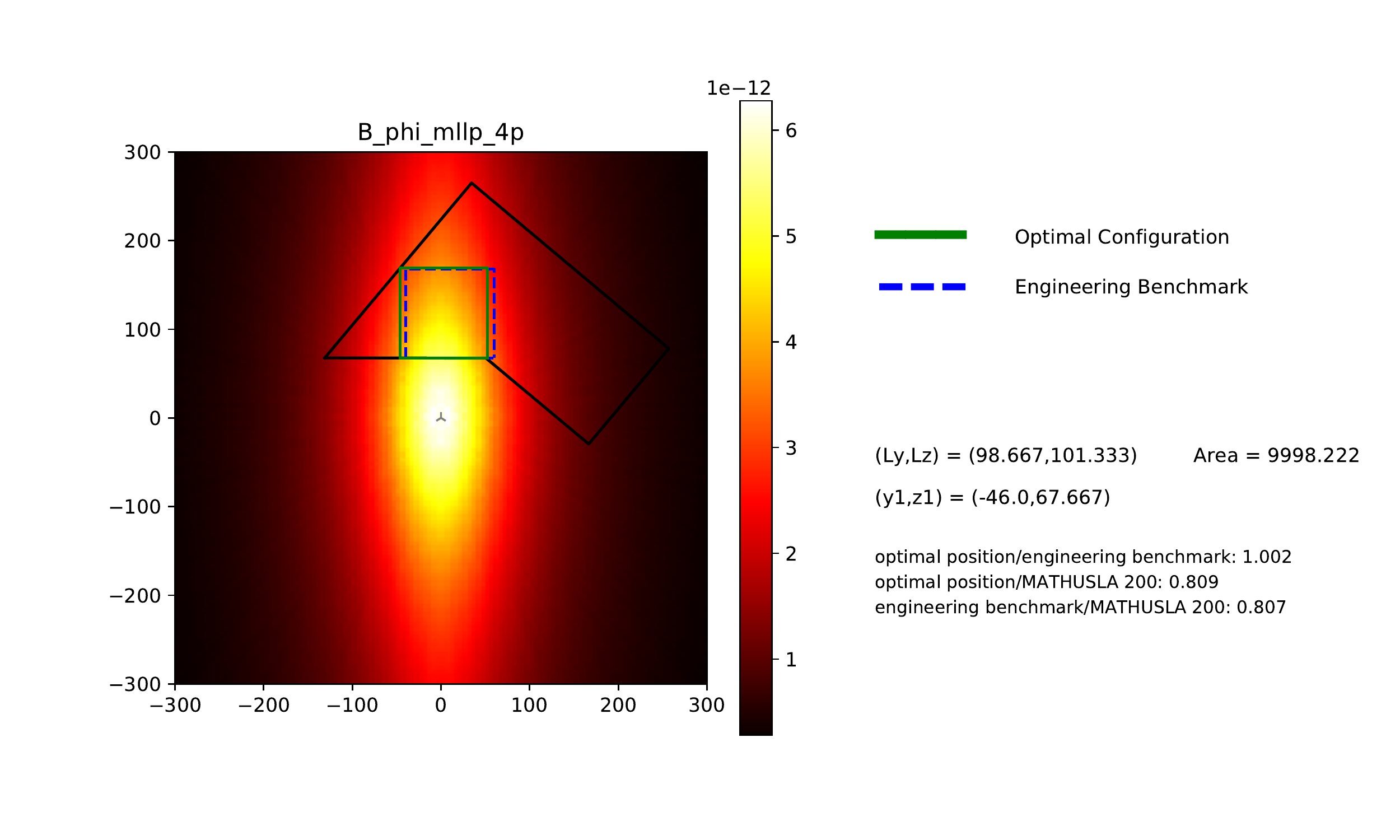}% Here is how to import EPS art
\caption{\label{fig:epsart}  p p $\rightarrow$ B $\rightarrow$ scalar, $m_{LLP} = 4.0$ GeV }
\end{figure}

\end{document}